\documentclass[12pt,preprint]{aastex}
\newcommand{\unitspace}{\ensuremath{\;}}
\newcommand{\usp}{\unitspace}
\newcommand{\unitstyle}[1]{\ensuremath{\mathrm{#1}}}

\newcommand{\kilo}{\unitstyle{k}}
\newcommand{\Mega}{\unitstyle{M}}

\newcommand{\meter}{\unitstyle{m}}
\newcommand{\cm}{\unitstyle{cm}}
\newcommand{\gram}{\unitstyle{g}}
\newcommand{\second}{\unitstyle{s}}
\newcommand{\K}{\unitstyle{K}}	           
\newcommand{\grampercc}{\gram\usp{\cm}^{-3}} 
\newcommand{\grampersquarecm}{\gram\usp{\cm}^{-2}} 
\newcommand{\GramPerCc}{\grampercc}
\newcommand{\GramPerSc}{\grampersquarecm}
\newcommand{\dyne}{\unitstyle{dyn}}
\newcommand{\erg}{\unitstyle{erg}}
\newcommand{\gauss}{\unitstyle{G}}
\newcommand{\ergpersecond}{\erg\unitspace{\second}^{-1}}
\newcommand{\eV}{\unitstyle{eV}}
\newcommand{\keV}{\kilo\eV}
\newcommand{\MeV}{\Mega\eV}
\newcommand{\Msun}{\ensuremath{M_\odot}}
\newcommand{\yr}{\unitstyle{yr}}     
\newcommand{\kB}{\ensuremath{k_\mathrm{B}}} 
\newcommand{\EF}{\ensuremath{\mathcal{E}_\mathrm{F}}} 
\newcommand{\NA}{\ensuremath{N_\mathrm{\!A}}} 
\newcommand{\ee}[1]{\ensuremath{\times 10^{#1}}}
\newcommand{\ddt}[1]{\frac{\partial #1}{\partial t}} 
\newcommand{\DDt}[1]{\frac{d #1}{dt}} 
\newcommand{\DDy}[1]{\frac{d #1}{dy}} 

\begin{document}

\title{A Remarkable Three Hour Thermonuclear Burst From 4U 1820-30}
\author{Tod E. Strohmayer}
\affil{Laboratory for High Energy Astrophysics, NASA's Goddard Space Flight 
Center, Greenbelt, MD 20771}
\email{stroh@clarence.gsfc.nasa.gov}
\and
\author{Edward F. Brown}
\affil{University of Chicago, Enrico Fermi Institute, 5640 South Ellis
Avenue, Chicago, IL 60637}
\email{brown@flash.uchicago.edu}

\begin{abstract}

We present a detailed observational and theoretical study of a $\sim
3$ hr long X-ray burst (the ``super burst'') observed by the Rossi
X-ray Timing Explorer (RXTE) from the low mass X-ray binary (LMXB) 4U
1820-30.  This is the longest X-ray burst ever observed from this
source, and perhaps one of the longest ever observed in great detail
from any source.  We show that the super burst is thermonuclear in
origin.  Its peak luminosity of $\sim 3.4 \times 10^{38}$ ergs
s$^{-1}$ is consistent with the helium Eddington limit for a neutron
star at $\sim 7$ kpc, as well as the peak luminosity of other,
shorter, thermonuclear bursts from the same source.  The super burst
begins in the decaying tail of a more typical ($\approx 20$ s
duration) thermonuclear burst.  These shorter, more frequent bursts
are well known helium flashes from this source.  The level of the
accretion driven flux as well as the observed energy release of
upwards of $1.5 \times 10^{42}$ ergs indicate that helium could not be
the energy source for the super burst.  We outline the physics
relevant to carbon production and burning on helium accreting neutron
stars and present calculations of the thermal evolution and stability
of a carbon layer and show that this process is the most likely
explanation for the super burst.  Ignition at the temperatures in the
deep carbon ``ocean'' requires $> 30$ times the mass of carbon
inferred from the observed burst energetics unless the He flash is
able to trigger a deflagration from a much smaller mass of carbon.  We
show, however, that for large columns of accreted carbon fuel, a
substantial fraction of the energy released in the carbon burning
layer is radiated away as neutrinos, and the heat that is conducted
from the burning layer in large part flows inward, only to be released
on timescales longer than the observed burst.  Thus the energy
released during the event possibly exceeds that observed in X-rays by
more than a factor of ten, making the scenario of burning a large mass
of carbon at great depths consistent with the observed fluence without
invoking any additional trigger.  A strong constraint on this scenario
is the recurrence time: to accrete an ignition column of $10^{13}$ g
cm$^{-2}$ takes $\sim 13 / (\dot M / 3 \times 10^{17}$ g s$^{-1}$) yr. 
Spectral analysis during the super burst reveals the presence of a
broad emission line between $5.8 - 6.4$ keV and an edge at $8 - 9$ keV
likely due to reflection of the burst flux from the inner accretion
disk in 4U 1820-30.  We believe this is the first time such a
signature has been unambiguously detected in the spectrum of an X-ray
burst.

\end{abstract}


\vfill\eject

\section{Introduction}

With an orbital period of only 11.4 minutes, 4U 1820-30 is the most
compact low mass X-ray binary (LMXB) known (Stella et al.  1987). 
Thermonuclear X-ray bursts were discovered from this source by
Grindlay \& Gursky (1976) and attest that the primary is a neutron
star.  Using Vela 5B data Priedhorsky \& Terrell (1984) found a 176
day periodic modulation of the X-ray flux from 4U 1820-30.  Bursts
have only been observed when the accretion driven flux is near the low
end of its observed range, $< 4\times 10^{-9}$ ergs cm$^{-2}$
s$^{-1}$.  Mass transfer in this system can be driven by gravitational
radiation at the rate of $\sim 3 \times 10^{17}$~g~s$^{-1}$.  The
binary resides in the globular cluster NGC 6624.  Optical observations
give a distance estimate of 7.6 kpc (Hesser \& Shawl 1985; Rich,
Minniti, \& Liebert 1993).  Based on the peak flux of photospheric
radius expansion X-ray bursts, Vacca, Lewin, \& van Paradijs (1986)
estimated a likely distance of $\sim 6.6$ kpc.  The compact nature of
the system requires that the secondary be a low mass helium dwarf (see
Rappaport et al.\ 1987), so that the accreted material likely has a
very high helium abundance.  The observation of photospheric radius
expansion bursts is consistent with the idea that helium flashes are
the primary fuel for the 10--20~s duration bursts most commonly
observed.  Modulations of the UV flux from 4U 1820-30 at the 11.4
minute orbital period were predicted by Arons \& King (1993) and
subsequently discovered by Anderson et al.\ (1997).  The UV
modulations result from X-ray heating of the secondary and constrain
the system inclination $i$ to the range $35^{\circ} \le i \le
50^{\circ}$.

Smale et al. (1997) reported the discovery of kilohertz quasiperiodic 
oscillations (QPO) from 4U 1820-30. Since then further observations have 
revealed that the highest observed kHz QPO appear to 
reach a saturation frequency near 1,050 Hz, suggesting the presence 
of a last stable circular orbit as predicted by General Relativity 
(GR) (Zhang, Strohmayer \& Swank 1998; Kaaret, Ford \& Chen 
1998). If this hypothesis is correct it also suggests that the neutron star 
in 4U 1820-30 may be as massive as $\sim 2 M_{\odot}$ (see also Arons \& 
King 1993). Constraints derived from the study of X-ray bursts can provide 
information on the neutron star mass and radius and thus might also provide a 
key test of these conclusions. 

Recently, Cornelisse et al.  (2000) reported the detection of a very
long ($\sim 3.5$ hr) burst from 4U 1735-44 with the Wide Field Camera
(WFC) on {\it BeppoSAX}.  Spectral softening in this burst led the
authors to conclude that it was thermonuclear in origin.  There have
been other recent reports of very long thermonuclear bursts from LMXBs,
for example, Heise, in 't Zand \& Kuulkers (2000) report long bursts
from KS 1731-260 and Serpens X-1, and Kuulkers (2001) recently identified a
likely super burst from GX 3+1 in RXTE/ASM data.  
As part of a Rossi X-ray Timing
Explorer (RXTE) AO3 observing program to try and detect X-ray bursts as
well as study kHz QPO from 4U 1820-30, RXTE was observing the source on
September 9, 1999 UT when a long, powerful X-ray burst was observed.
This burst was $\sim 1,000$ times longer than the more frequent
thermonuclear bursts seen from the source (see for example Haberl et al.
1987).  In this paper, we show that this super burst was thermonuclear
in origin and that it was most likely fueled by burning of carbon ashes
produced by the stable burning of helium accreted onto the neutron star
in 4U 1820-30.  In this respect, 4U~1820-30 differs from the other five
super burst sources, which accrete a hydrogen/helium mixture and
therefore produce much less carbon (see Cumming \& Bildsten 2001).  A
consequence of this is that the recurrence time for this
source will be of the order of a decade, however, unless the carbon
ignition is somehow ``triggered.''

A very brief summary of the analysis in this paper was previously
presented by Strohmayer (2000).  We also note that during the
preparation of this paper a burst lasting several hours was observed
during RXTE observations of 4U 1636-53 in support of RXTE proposal
50030 (PI: Strohmayer).  An analysis of this burst will be presented
elsewhere.  The 4U 1636-53 event was independently found with the ASM by 
Wijnands (2001), who also found a similar event in the ASM archive which 
occurred $\sim 5$ years earlier.

The plan of this paper is as follows.  In \S~2 we present the observed
temporal and spectral characteristics as well as the energetics of the
super burst.  In \S~3 we discuss the implications of the discrete
spectral components (line and edge) observed during the burst. 
Section~4 begins with a discussion of the production of carbon and the
thermal structure of the ``ocean.''  We assess the amount of carbon
needed to produce a thermonuclear runaway and the evolution of the
ensuing burst.  We conclude in \S~5 with a summary of our principal
findings.

\section{The X-ray Super Burst: Observations}

As mentioned above, 4U 1820-30 has a $\sim 178$ day X-ray intensity
modulation (Priedhorsky \& Terrell 1984).  The All-sky Monitor (ASM)
on board RXTE has been monitoring the source for more than 5 years. 
Figure 1 shows a portion of the 2--12 keV ASM lightcurve of 4U
1820-30.  The dips down into the lowest flux state occur approximately
at the 178 day period.  Our RXTE observations on MJD 51430 (Sep.~9,
1999 UTC, indicated by the vertical arrow in the figure) caught
the source near its low flux state when bursts can be observed.

In the weeks prior to our observation RXTE had a malfunction in one of
its antenna transponders which disabled one of the high-gain antennas
with which RXTE telemeters data to the ground.  Because of this
antenna malfunction some data from our observation were unfortunately
lost.  For the Proportional Counter Array (PCA) we have only Standard1
and Standard2 data for the entire event.  These data provide 2--90
keV time histories with 1/8 s temporal resolution (Standard1) and 16 s
spectral accumulations across the 2--90 keV bandpass (Standard2),
respectively. Although the PCA bandpass is nominally 2 - 90 keV, for these
data almost all source counts are in the 2 - 40 keV band. 
We did obtain some high time resolution event mode data
(sampling rate of 1/8192 Hz) but only for two time intervals in the
decaying portion of the burst.  Unfortunately, we lost the high time
resolution data during the onset of the burst.  We will investigate
the timing properties of 4U 1820-30 at the time of the super burst in
a subsequent publication.

\subsection{General Description}

The time history of the burst is shown in figure 2.  The 2--90 keV PCA
Standard1 lightcurve with 1/8 s time resolution is the higher time
resolution histogram (left axis), while the (8--30)/(2--8)~keV
hardness ratio from Standard2 data is the lower resolution histogram (right
axis). The burst was observed with 3 of the 5 proportional counter units
(PCUs 0, 2, and 3) which comprise the PCA. Note that the time axis is
logarithmic.  The event began with a typical helium flash from 4U
1820-30 which extends from about 3--15 seconds in Figure 2.  However,
before this first burst dies away the super burst erupts and is still
decaying away 10,000 s later when our observation ended.  The event
shows a clear `precursor' after which the X-ray flux dropped
completely to background for a few seconds.  A similar, but much
weaker precursor can be seen in the helium flash burst just prior to
the onset of the super burst.  This behavior is characteristic of very
powerful photospheric radius expansion bursts (see Hoffman et al. 
1978; Tawara et al.  1984; Lewin, Vacca \& Basinska 1984; van Paradijs
et al.  1990).  Note also the hardening of the spectrum following the
precursor.  This is characteristic of the contraction phase of a
radius expansion burst.  We found in the RXTE archive one additional
example of a helium flash burst from 4U 1820-30.  In figure 3 we
compare this burst, observed on May 2, 1997 at 17:32:39 UTC, with the
helium flash which preceded the super burst.  This He burst
was observed with all 5 PCUs, so in order to make a closer comparison
we have scaled the observed count rate by 3/5.  Figure 3 shows the
lightcurve of each burst at 1/8 s resolution from the Standard1 data
mode.  There are several things of note: both the preburst count rates
as well as the peak count rate during the burst are similar, and each
burst has a precursor or `double peak' which indicates photospheric
radius expansion.  The above features of the time profile, combined
with the observed behavior of the hardness ratio, are consistent with
a thermonuclear origin for both bursts.  The much longer timescale of
the super burst compared to the shorter, more frequent helium flashes
suggests a larger reservoir of nuclear fuel released at much greater
column depth below the surface.  These considerations suggest that
carbon and/or oxygen are the likely fuel sources for the super burst. 
Later, in \S~4, we will explore this scenario in greater detail.

\subsection{Spectral Analysis}

We used the Standard2 data to investigate the X-ray spectrum both before
and throughout the burst. For most of the burst we did not have spectral
data in a high time resolution mode. This was unfortunate, but since
the evolution timescale of this burst is much longer than a typical 10 s
burst it was only a significant problem near the onset of the burst
where the data suggest the black body temperature was changing on a
timescale shorter than our 16 s accumulations.

Before investigating the burst spectrum we analyzed the spectrum of
the persistent, accretion driven flux.  We used the standard PCA
background estimation software, PCABACKEST, to determine the PCA
detector background.  For all the spectral results we used only the
top layer of detectors 0, 2, and 3 of the PCA. We found that the
persistent flux before the burst is well fit by a thermal
bremsstrahlung model (bremss in XSPEC) with a temperature $kT_{brem}
\sim 8$ keV and flux of $3.5\times 10^{-9}$ ergs cm$^{-2}$ s$^{-1}$. 
Figure 4 shows the time history of the thermal bremsstrahlung
temperature, $kT_{brem}$ (top panel), and the derived 2--20 keV flux
(bottom panel).  At a distance of 6.6 kpc the observed flux corresponds
to a luminosity of $1.8 \times 10^{37}$, consistent with previous
measurements of the persistent luminosity when bursts are often
observed (Clark et al.  1977; Haberl et al.\ 1987).

In most published analyses of burst spectral evolution the spectrum 
is obtained by subtracting off an estimate of the spectrum of the preburst,
accretion driven flux. This should be reasonable when the burst flux does not
appreciably disturb the accretion flow, however, during powerful radius 
expansion bursts it may be that the burst driven wind and near-Eddington flux 
can substantially effect the accretion process (see, for example, Walker \&
Meszaros 1989). We investigated both subtracting off 
the preburst emission as a background and assuming that the accretion-driven 
flux is completely shut off by the burst. For the latter case we simply 
subtracted off the appropriate detector background estimated with PCABACKEST. 
In general we found that for the majority of the burst we were able to get 
better fits by subtracting off the preburst spectrum, so we will discuss our 
results using this method. The best fit parameter values showed modest 
quantitative differences ($< 10 \%$) when estimated using the two different 
background assumptions, but all of our qualitative conclusions are robust in 
the sense that they were independent of which assumption we used for the 
background. 

Because this burst is $\sim 1,000$ times longer than typical thermonuclear
bursts the timescale on which the thermal continuum changes is also much 
longer. This makes it possible to obtain very high signal to noise spectra 
during intervals in which the black body temperature remains fairly 
constant. For example we were able to accumulate 64 s spectra throughout most 
of the burst and still obtain good fits to the continuum with single 
temperature black body models. The spectrum during the burst is indeed thermal,
and evolves in a way which is entirely consistent with thermonuclear bursts 
of shorter duration. 

We found that a black body model with photoelectric absorption (model
wabs*bbodyrad in XSPEC) alone usually did not provide an acceptable fit
in the statistical sense, however, non-thermal continua gave much worse
fits.  The residuals when fitting a black body function and
photoelectric absorption strongly suggest the presence of an emission
line near 6.4 keV as well as an absorption edge between 7--9 keV. To
model these components we included a gaussian emission line as well as
an edge in our spectral model (model wabs*(bbodyrad+gaus)*edge in
XSPEC). With these components included we were able to obtain acceptable
fits with $\chi^2$ per degree of freedom $\sim 1$. Figure 5 shows the
characteristic residuals near the peak of the burst when the gaussian
line and edge are removed from the model. Shown are the residuals, data
minus model, in units of standard deviations. These residuals appear to
be qualitatively similar to those reported by van Paradijs et al. (1990)
for a powerful burst from 4U 2127+11 in M15. Both the line and edge are
strongly required to adequately model the data. Figure 6 shows the count
rate spectrum and residuals from a typical fit near the peak of the
burst including the line and edge components, and demonstrates the
quality of fit we can achieve with this model.  In general we find that
the gaussian line at $\sim 6.4$ keV has finite width and often an energy
centroid significantly less than 6.4 keV, suggestive of a line produced
by reflection from a relativistic accretion disk in 4U 1820-30.  Day \&
Done (1991) had suggested that such a signature should be present in
burst spectra, but to our knowledge this is the first compelling
observational evidence for such an effect, and it opens up the prospect
of probing directly the properties of accretion disks during long
thermonuclear bursts. With the disk reflection mechanism in mind we also
modeled the line as an Fe fluorescence feature from an accretion disk
(model diskline in XSPEC). Since the absorption edge will also be
broadened by motions in the disk, we also included a smeared edge in
these fits (model smedge in XSPEC).
 
We obtained excellent fits with this model for most of the burst duration
except for about 100 s after the precursor.  During this interval the 
photosphere is extended and the black body temperature is evolving too quickly 
for the 16 s Standard2 accumulations to resolve. For this interval we would 
have greatly benefited from higher time resolution spectra.  
We summarize our investigation of
the time evolution of the burst spectrum in figures 7, 8 and 9.  Figure
7 shows the time history of the burst from the Standard1 data on a
linear time axis.  The vertical dashed lines denote the region in which
we derived black body fits.  The time evolution of the spectral
parameters is shown in figures 8 and 9.  In each case the timespan is
that of the interval between the dashed vertical lines in figure 7.  The
derived black body flux, temperature, and inferred radius at 6.6 kpc are
shown in figure 8.  The evolution of the line and edge parameters is
shown in figure 9.  The evolution of the black body temperature and flux
is very similar to that of other thermonuclear bursts, the only
difference being that the timescale of the evolution is much longer.
For comparison we also investigated the spectral evolution during the
shorter, helium flash observed from 4U 1820-30 on May 2, 1997.  The peak
black body temperature of both bursts approaches $\sim 3$ keV, and the
peak fluxes are also consistent.  This leaves no doubt about the
thermonuclear origin of the super burst.

The super burst from 4U 1820-30 provides a unique opportunity to try and use 
burst spectroscopy to constrain the radius of a neutron star. The fact that 
the source distance is reasonably well constrained, the accreted matter is 
likely pure helium, and the burst lasted so long--giving high signal to noise 
spectra--make such an effort well worthwhile.  Our aim in this paper is to lay 
out the basic properties of the super burst and a likely mechanism for its 
production. It is beyond the scope of this paper to present a detailed
spectroscopic study in an attempt to derive constraints on the neutron star 
radius, however, we can make some simple estimates using our black body 
spectral fits and previous calculations of pure helium, neutron star 
atmospheres. The spectrum of an X-ray burst is not expected to be strictly 
that of a black body. Many authors have investigated the effects of electron 
scattering on the spectrum of X-ray bursting atmospheres, which is the dominant
opacity source under the conditions typical during bursts (see for example, 
London, Taam \& Howard 1986; Ebisuzaki 1987; Titarchuck 1994; Madej 1991; 
Madej 1997). What
is found is that the observed color temperature (ie. the temperature derived by
fitting a black body function) is higher (harder) than the actual effective 
temperature of the atmosphere by a factor which ranges from about 1.4 to 1.7 
and is largely dependent on the local X-ray flux. This correction factor is 
usually referred to as the hardening factor. For example, Ebisuzaki \& Nakamura
(1988) provide analytic approximations for the hardening factors in X-ray
bursting, pure helium atmospheres. These factors depend only on the ratio
of the local flux to the Eddington flux.  Using these hardening factors we
can convert our measured color temperatures to effective temperatures and 
then infer the size of the neutron star using the formula;
\begin{equation}
f_{\infty} = R^2 (1 - 2GM/c^2R)^{-1} \sigma_R T_{eff}^4 / d^2  \; ,
\end{equation}
where $f_{\infty}$, $T_{eff}$, $\sigma_R$, and $d$ are the flux, effective 
temperature (both measured far from the neutron star), Stefan - 
Boltzmann constant, and the source distance,  respectively. Here $M$ and $R$ 
are the gravitational mass and radius of the star, both measured at the stellar
surface. Because of the general relativistic effects this expression does not 
directly give an estimate of the radius independent of the mass. We used the 
flux and temperature measurements summarized in Figure 8 along with the 
hardening factors from Ebisuzaki \& Nakamura (1988) to evaluate the above 
expression (assuming a distance of 6.6 kpc). For each pair of 
flux - temperature measurements we get an estimate of the quantity,
\begin{equation}
g(M, R) = R (1 - 2GM/c^2R)^{-1/2} \; .
\end{equation}
The mean value of $g(M,R)$ in these data is 18.3 km, which implies a mean 
radius range of 15.7 to 13.85 km for a mass range of 1.4 to 2 $M_{\odot}$. 
This range of radii is about a factor of two larger than that determined by 
Haberl \& Titarchuk (1995) for 4U 1820-30.  We will explore the implications 
of the super burst for the mass - radius relation in 4U 1820-30 in greater 
detail in a sequel. 

\subsection{Observed Burst Energetics}

The peak flux of $6.5 \times 10^{-8}$ ergs cm$^{-2}$ s$^{-1}$ during the
super burst implies a peak isotropic luminosity of $3.4 \times 10^{38}$
ergs s$^{-1}$ at a distance of 6.6 kpc.  A neutron star with a pure
helium photosphere has an Eddington luminosity of $L_\mathrm{Edd} =
2.5\times 10^{38} (M/M_{\odot})(1 -
2.9644(M/M_{\odot})/R_\mathrm{km})^{1/2}$ ergs s$^{-1}$, which for
reasonable neutron star masses and radii is consistent with our inferred
peak luminosity.  To determine the total energy in the observable burst
we integrated the flux vs time profile, linearly interpolating across
the gaps.  We find a total fluence of $2.7 \times 10^{-4}$ ergs
cm$^{-2}$.  At a distance of 6.6 kpc this implies a total energy release
of $1.4 \times 10^{42}$ ergs.  We note that these numbers should be
considered lower limits since when our observations ended the burst flux
was still decaying away.  Moreover, it is likely that a substantial
fraction of the energy released in the carbon burning layer is emitted
as neutrinos or conducted inward and only released on a timescale much
longer than the observed burst (see our discussion in
\S~\ref{subsec:thermal-decay}).  Nevertheless, the observed X-ray
fluence is $\sim 1,000$ times greater than is released in a typical
helium flash from 4U 1820-30 (Haberl et al.  1987).  The basic physics
of nuclear burning instabilities on neutron stars indicates that carbon
and oxygen burning can in principle provide this amount of energy, but
not helium (see for example Lamb \& Lamb 1978).

\section{Implications of the Spectral Line and Edge}

Franco \& Strohmayer (1999) reported evidence for a $\sim 6.4$ keV
emission line during the expansion phase of the May 2, 1997 burst from
4U 1820-30 observed with RXTE. There have been several previous claims
of absorption lines between 4 and 5 keV from several burst sources, but
their reality and interpretation have been controversial (see for example 
Waki et al.\ 1984; Nakamura, Inoue, \& Tanaka 1988; Magnier et al.\ 1989;
Foster, Ross, \& Fabian 1987; Madej 1989), and we note that these previous 
line claims have not been confirmed by subsequent observations, for example,
with the Advanced Satellite for Cosmology and Astrophysics (ASCA). 

Our results leave no doubt
that emission lines consistent with Fe K$\alpha$ fluorescence can be
produced during X-ray bursts.  What is the origin of these discrete
spectral features?  A number of authors have investigated line
production mechanisms during X-ray bursts.  Foster, Ross, \& Fabian
(1987) investigated the formation of Fe features in the spectrum of
cooling X-ray bursts.  Their models predict broad K$\alpha$ emission
features should be present along with a `blend' of photoelectric edges
near 9.1 keV.  We see both a broad line and an edge during the super
burst, features similar to those predicted by Foster, Ross, \& Fabian
(1987), however, the observed energies of the features are somewhat
problematic for this interpretation because the inferred redshifts are
uncomfortably small for what would be expected from a canonical neutron
star.  For example, when our observed black body temperature reaches a
maximum, the inferred line centroid is about 6.2 keV (see figures 8 and
9).  If this resulted from a redshift of a 6.9 keV line of hydrogen-like
Fe, one would require a redshift $1+z = (1 -
2.9644(M/M_{\odot})/R_\mathrm{km})^{-1/2} \approx 6.9/6.2 = 1.113$ which
would imply a neutron star of radius 21.5 km for a mass of $1.4
M_{\odot}$.  This is prohibitively large even for the stiffest estimated
neutron star equations of state.  One might argue that the neutron star
photosphere is still extended, but the temperature is thought to peak
when touchdown of the photosphere has occurred.  If the rest energy of
the line were 6.7 keV then the problem is only exacerbated. It is possible,
however, that because of the modest spectral resolution of these observations,
the measured line centroid could be influenced by the choice of the continuum 
model, so these conclusions should still be considered cautiously.

As mentioned previously, Day \& Done (1991) suggested that a disk
reflection component could be detected in the spectrum of X-ray
bursts, and that detection of the absorption edge could be used to
probe the ionization state of the disk.  We think this is the simplest
interpretation of the discrete features we see in the burst spectra. 
The inferred line centroids, which range from $\sim
5.8\mbox{--}6.4$~keV are comfortably consistent with an origin in the
inner accretion disk.  Relativistic motions in the disk also broaden
the line profile.  Furthermore, the inferred edge energies of $\sim
7.5\mbox{--}9$~keV are also consistent with this picture.  Changes in
the inferred energy of the absorption edge also suggest corresponding
changes in the ionization state of the disk.  For example, the highest
inferred edge energies correspond to the time when the black body flux
and temperature were near their peaks, suggesting that the burst flux
has a significant effect on the ionization state of the disk.  The equivalent
width of the $\sim 6.4$ keV line ranges from $\sim 70 - 200$ eV, but is
typically about 130 eV.  This is reasonably consistent with recent estimates
derived from spectral modelling of reflection from X-ray illuminated disks
(see for example, Li, Gu \& Kahn 2001; Nayakshin \& Kallman 2001; Ross, Fabian
\& Young 1999). In principle, if the energies of the emission line and edge
are known then one can constrain the ionization state of the gas.  Turner 
et al. (1992) tabulate the run of Fe K line and edge energies with ionization
parameter in the context of Fe K$\alpha$ observations of Seyfert galaxies (see
their Figure 2). Based on this compilation, our line and edge energies appear 
consistent with an ionization parameter $\sim 100$ near the peak of the super
burst. The subsequent evolution of the edge and line energies through the 
burst appears consistent with a decreasing ionization parameter as the burst 
evolves. This seems sensible, since the ionizing flux (from the burst) is 
decreasing with time. However, more precise spectral modelling will be required
to investigate the implications for disk structure in detail. This will be left
for a future study. 

During the decay of the burst there is a highly significant hardening
of the spectrum (see figure 8, near 3000 s).  This is evidenced by an
increase in the black body temperature along with a decrease in the
inferred black body radius.  Even more interesting, the line flux and
edge depth both decrease dramatically at the time of this spectral
hardening (see figure 9).  This behavior is somewhat reminiscent of
state changes in other accreting X-ray binaries, but happens on a much
shorter timescale.  Perhaps the inner disk is transitioning to a hot,
optically thin condition, similar to an ADAF (see for example Narayan,
Yi, \& Esin 1998).  Alternatively, we could be seeing a geometric
effect of something in the system shadowing the disk so that the
reflected component is greatly reduced, perhaps a warp in the
accretion disk.  Unfortunately, an Earth occultation interval blocks
our view of what happens next.  Also of note is the fact that shortly
after this hardening episode chaotic dipping can be seen in the
lightcurve for an extended period.  Although for part of this time the
source is Earth occulted, this dipping may be another indication of
the accretion flow having been strongly affected by the burst.

Another possibility for the discrete components is that we are seeing
fluorescence in the wind material blown off the star by the
super-Eddington burst. In this case one might expect to see a
correlation of the line strength with the amount of absorbing material
along the line of sight, and this does not seem to be the case.

It seems likely that at some level all of these processes are at work,
but it seems most plausible that disk reflection is the dominant
mechanism. The UV modulations observed by Anderson et al.\ (1997)
suggest the system has an inclination $\sim 45^{\circ}$ (see Arons \&
King 1993), which would provide a favorable geometry for reflection. The
energy resolution of the PCA is not sufficient to resolve the line and
edge in detail. For example, a redshifted line from the neutron star
photosphere could be obscured by the broad line produced by disk
reflection. However, additional observations of such a burst with much
higher spectral resolution could help distinguish amongst these
different scenarios, and could provide us with a powerful new probe of
the accretion process in LMXBs.

\section{The Case for Unstable Carbon Burning}\label{sec:theory}

Having described the observation of the super burst and the evidence for
it being thermonuclear in origin, we now examine its source of fuel and
its ignition.  In order for He burning to power the super burst, the
accumulated He must not ignite until the base of the helium layer
reaches densities of order $10^{8}\usp\GramPerCc$
\citep[e.g.,][]{fryxell.woosley:two-dimensional,zingale.timmes.ea:helium}.
This can only occur if the atmosphere is extremely cold and in
particular requires a local accretion rate $\lesssim
2\ee{-11}\usp\Msun\usp\yr^{-1}$ \citep{fryxell.woosley:two-dimensional}
in the absence of any heat flux from the core.  Such low accretion rates
are not consistent with the observed flux and the presence of concurrent
He flashes.  The fact that a helium flash went off just prior to the
super burst also indicates that helium cannot be the primary energy
source for the giant burst.  We therefore discount the possibility that
$\mathrm{^{4}He}$ could be the fuel for the super burst.

As shown in the previous section, the spectral evolution of the burst
is roughly consistent with the release of energy at large densities in
the star.  The burning of carbon to iron-peak elements releases
$\approx 10^{18}\usp\erg\usp\gram^{-1}$ so that at least $ 2 \times
10^{24}\usp\gram$ of carbon is needed to account for the energy
released in the burst (in \S~\ref{subsec:thermal-decay} we argue that
this is likely an underestimate of the total burst energetics).  An
$M_\mathrm{C} = 2\ee{24}\usp\gram$ mass of $\mathrm{^{12}C}$ uniformly
deposited over the surface of a neutron star of radius $R =
10\usp\kilo\meter$ has a column density $y_\mathrm{C} \approx
M_\mathrm{C}/4\pi R^{2}\approx 1.6\ee{11}\usp\GramPerSc$.  Hydrostatic
balance gives the pressure at the base of the layer, $p = g y$; from
the equation of state and the thermal conductivity (see
\S~\ref{subsec:thermal-decay} for details) we may estimate the cooling
timescale for a mass of newly formed $\mathrm{^{56}Fe}$ to be $\sim
C_{p}y^{2}/(\kappa\rho) \sim
10^{3}(y_\mathrm{C}/10^{11}\usp\GramPerSc)^{3/4}\usp\second$, roughly
consistent with the observed thermal decay timescale of the burst. 
Here $C_p$ is the specific heat, per unit mass, $\kappa$ is the
thermal conductivity, $\rho$ is the density, and $y = \int_z^\infty
\rho\,dz'$ is the column depth.  If mass transfer is driven by
gravitational radiation we expect an accretion rate of $\sim
1\mbox{--}3 \times 10^{17}\usp\gram\usp\second^{-1} \sim 5 \times
10^{-9} M_{\odot}\usp\yr^{-1}$ onto the neutron star; this level is
also roughly consistent with the accretion rate required to account
for the persistent luminosity of $1.8 \times
10^{37}\usp\ergpersecond$.  The precise rate is uncertain because of
our lack of knowledge concerning the efficiency with which
gravitational energy is converted to X-ray flux.  If this conversion
were 100\% efficient, then one would require $\dot{M} =
10^{17}\usp\gram\usp\second^{-1}$ (assuming a 1.4 $M_{\odot}$ and
$10\usp\kilo\meter$ neutron star); a more realistic number, however,
is probably 5--20 \%, so that the actual mass accretion rate could be
higher than $10^{18}\usp\gram\usp\second^{-1}$.

Isotropic (over the stellar surface) accretion at a rate of $3 \times 10^{17}$
g s$^{-1}$ can deposit a sufficient amount of carbon to power the burst only 
after 0.2--0.6\usp\yr, so such events should indeed be rare, consistent with
only one such burst being observed from this source.

Unstable carbon burning on accreting neutron stars was conjectured by
\citet{woosley.taam:carbon} and studied by \citet{taam78:_nuclear},
who found that for accretion rates of
$10^{-10}$--$10^{-9}\usp\Msun\usp\yr^{-1}$, the timescale for
recurrence would be of order 10--100\usp\yr, with the bursts having an
energy of order $10^{44}\usp\erg$, about 100 times greater than that
measured for this burst.  While the original calculation of
\citet{taam78:_nuclear} assumed accretion of hydrogen-rich material,
recent calculations with large reaction networks for both unstable
\citep{schatz.aprahamian.ea:endpoint} and steady-state burning
\citep{schatz99} find that only a small amount of $\mathrm{^{12}C}$ is
produced during hydrogen burning.  For this reason, the scenario
presented here (i.e., the ignition of a carbon-rich layer) is probably
not relevant for the super bursts seen from 4U~1735--44, KS 1731-260,
GX~3+1, Ser X-1 and 4U 1636-53, which are likely accreting a mix of
hydrogen and helium (see \citealt{cornelisse.ea:longest}; Heise, in 't
Zand, \& Kuulkers 2000; \citealt{kuulkers:superoutburst}; Wijnands
2001).  The small amount of $\mathrm{^{12}C}$ produced during
rp-process burning might be a fuel for the other super-bursts; for a
discussion of this possibility, we refer the interested reader to
\citet{cumming.bildsten:carbon}.

While the burning of $\mathrm{^{12}C}$ is a plausible cause of this
super burst, the ignition of a pure carbon layer requires either high
temperatures \citep{brown98} or a large mass of carbon
\citep{taam78:_nuclear}.  In this section, we address these
constraints.  We begin (\S~\ref{subsec:C12-production}) by estimating
the amount of carbon produced.  We then
(\S~\ref{subsec:ignition-carbon-layer}) investigate the conditions for
ignition of a carbon-rich layer.  We demonstrate, in the absence of a
``trigger,'' that a much larger mass of $\mathrm{^{12}C}$ is required
than that estimated from the observed burst energetics. 
Section~\ref{subsec:thermal-decay} provides a resolution to this
dichotomy: namely, for very large bursts most of the heat is carried
off by neutrinos, and of the heat conducted from the burning layer, a
substantial amount flows \emph{inward}, to be released on much longer
timescales (days to weeks).  The long timescale for thermal diffusion
at great depths means that only a fraction of the total burst energy
is radiated away in the few hours following the burst rise.  A
consequence of requiring a massive carbon layer for ignition is that
the recurrence times of such super bursts will be many years.

\subsection{The Production of $\mathrm{^{12}C}$}
\label{subsec:C12-production}

Whether or not $\mathrm{^{12}C}$ is produced by the burning of
$\mathrm{^{4}He}$ depends largely on whether the burning is stable.  The
ashes of stable He burning are composed mostly of $\mathrm{^{12}C}$ at
the lowest accretion rates compatible with stability \citep{brown98}.
In contrast, unstable burning leads to much higher temperatures, which
allow the reactions to run up to iron-peak nuclei.  Numerical
calculations find that for unstable burning only a trace amount of
residual He is expected ($< 1\%$ by mass; \citealt{joss80:_helium}) and
no $\mathrm{^{12}C}$.  Because this source accretes nearly pure He,
the endpoint of the unstable nucleosynthesis will be iron-peak nuclei,
rather than the heavier species produced during mixed H/He burning.

The luminosity of this source is variable by a factor of $\approx 4$
(Figure~\ref{fig1}), and type I X-ray bursts are only observed when the
luminosity is in its low state.  The lack of unstable burning when the
luminosity is high is somewhat puzzling, since the inferred accretion
rate is still less than what is necessary for stability, according to
linear analysis and time-dependent calculations \citep[see][and
references therein]{bildsten:thermonuclear}.  One possibility is that
the burning is unstable, but that the flame front propagates at the slow
conductive velocity \citep{bildsten95:_propag}.  Whether such fronts can
slowly propagate on a rapidly rotating neutron star is an open question
\citep[for a discussion, see][]{spitkovsky.ea.propagation}.  Another
possibility is that the accretion flow does not spread over the entire
surface prior to He ignition, as suggested by \citet{bildsten.theory} as
an explanation for patterns of unstable hydrogen/helium burning on
Z-sources. If the efficiency of conversion of gravitational energy to X-ray
luminosity is modest, then the inferred accretion rate can be 
substantially higher, reducing the discrepancy between theory and observation. 

One constraint on the amount of carbon produced would be the
recurrence time of a super burst.  We did search the RXTE/ASM archival
lightcurve of 4U 1820-30 for additional instances of super bursts, but
did not find any.  Although the ASM monitors the whole sky it does not
do so all the time, so that the total time during which another super
burst from 4U 1820-30 could have been seen is only about 25 days out
of a time span of about 5 years.  Since only one such event was seen
and the total on source time was modest we can not place a strong
constraint on the recurrence rate.  It is unlikely to be much larger
than 0.05 day$^{-1}$, but could easily be ten times lower than this. 
We note that the source is only at or below the ASM flux level at
which thermonuclear bursts were observed for about $10 \%$ of the
time.  For the purposes of this paper, we take the absence of observed
type~I X-ray bursts during the episodes of high-luminosity as
indication that $\mathrm{^{12}C}$ is produced.  Over each 178 day
cycle, we estimate that roughly $(1/2)\times 178\usp\mathrm{d}\times
\dot{M} \approx 10^{24}\usp\gram$ of $\mathrm{^{12}C}$ is formed.  As
we describe in the next section, it is quite plausible that the iron
made in the bursts mixes with the carbon made during stable burning,
so that the deep ocean is a mixture of the two.

\subsection{Ignition of the Carbon Layer}
\label{subsec:ignition-carbon-layer}

The physics involved in the ignition of the carbon layer is detailed in
\citet{brown98}; here we review that discussion.  The scale height is of
order $1000(y/10^{11}\usp\GramPerSc)^{1/4}\usp\cm\ll R$, so that the
carbon layer is susceptible to a thin-shell instability
\citep{hansen75:_thin}, and we may neglect the variation of gravity with
depth.  

We compute the thermal structure of the carbon layer following the
method outlined in \citet{brown98}.  The relevant equations are those of
continuity (for each species),
\begin{mathletters}
\begin{equation}\label{e:cont}
   \ddt{X_j} + \mathbf{v\cdot\nabla}{X_j} = 
	\frac{A_j}{\NA\rho} (-r_{\rm dest}^{(j)} + r_{\rm prod}^{(j)}),
\end{equation}
entropy,
\begin{equation}\label{e:entra}
   T\DDt{s} = -\frac{1}{\rho}\mathbf{\nabla\cdot F} + \varepsilon,
\end{equation}
and flux,
\begin{equation} \label{e:fluxa}
   \mathbf{F} = -\kappa\mathbf{\nabla} T.
\end{equation}
\end{mathletters}
Here $X_j$ and $A_{j}$ are the mass fraction and mass number of
species $j$, $\NA=6.02\ee{23}\usp\gram^{-1}$ is Avogadro's number, and
$r_{\rm dest}^{(j)}$ and $r_{\rm prod}^{(j)}$ are the volumetric
destruction and production rates of species $j$ via nuclear reactions.

For the electron equation of state, we use a table interpolation
scheme of the Helmholtz free energy \citep{timmes.swesty:accuracy}. 
The ionic equation of state includes Coulomb interactions, which are
parameterized by $\Gamma\equiv (\langle Z \rangle e)^{2}/(a \kB T)$,
where $4\pi a^{3}/3 = n_{I}^{-1}$ is the mean volume per ion.  In this
problem, $\Gamma$ spans a range of values $\ll 1$ (ions are weakly
coupled) to $> 173$ (ions are crystalline; see \citealt{farouki93} and
references therein).  For $1<\Gamma\lesssim 173$ the ions are in a
liquid state, and we shall refer to this layer as the ``ocean''.

The thermal conductivity $\kappa$ contains contributions from both
radiative transport (in the diffusive approximation), and electronic
thermal conduction.  Radiative transport dominates at lower densities;
in this regime we use the Rosseland mean of both Thompson scattering
(with corrections for the relativistic, partially degenerate
electrons; \citealt{sampson:opacity,buchler.yueh:compton}) and
free-free absorption with an analytical approximation to the Elwert
factor \citep{schatz99}, which reproduces the tabulated gaunt factor
of \citet{itoh91:_rossel_gaunt}.  At higher densities electron
conduction dominates; we include contributions from electron-electron
scattering, for which we use the fit \citep{potekhin97} to the
calculations of \citet{urpin80:_therm}, and electron-ion scattering,
for which we use the fits of \citet{potekhin99:_trans}.  The treatment
of \citet{potekhin99:_trans} is valid in both the liquid and
crystalline phases.

We are interested in the unstable ignition of $\mathrm{^{12}C}$.  Prior
to ignition, the heating from $\mathrm{^{12}C}+\mathrm{^{12}C}$ is
negligible, and we do not include it in the thermal structure
calculation.  In the carbon ocean and iron crust, then, $\varepsilon =
-\varepsilon_{\nu}$, the neutrino emissivity, for which we use the fits
of \citet{itoh96:_neutr} to the pair, plasma, photoneutrino,
bremsstrahlung, and recombination rates.  We change the composition
discontinuously from a mixture of $\mathrm{^{12}C}$ and
$\mathrm{^{56}Fe}$ (we allow for the possibility of a mixed layer) to
$\mathrm{^{56}Fe}$ at an interface column $y_{i}$, which is chosen to be
roughly where we expect the $\mathrm{^{12}C}$ to have ignited.

The thermal timescale, at a given column, is much faster than the
timescale for accretion to supply that column of material.  As a result,
we may simplify our equations by presuming steady-state ($\partial_t \to
0$) and converting the structure equations
(eq.~[\ref{e:cont}]--[\ref{e:fluxa}]) from PDE's into ODE's.  To further
simplify calculations, we rewrite the equations with the column $y$ as
the independent variable, $\partial_z = -\rho\partial_y$, and neglect
differential sedimentation of the ions.  With these approximations, the
velocity of the fluid is $\mathbf{v} = -\mathbf{e_{z}}\dot{m}/\rho$,
where $\dot{m}$ is the accretion rate per unit area, and the structure
equations reduce to
\begin{mathletters}
    \begin{eqnarray}
	\label{eq:He}
	\DDy{X_\mathrm{He}} &=& -\lambda_{3\alpha} \dot{m}^{-1}\\
	\label{eq:C}
	\DDy{X_\mathrm{C}} &=& \lambda_{3\alpha} \dot{m}^{-1}\\
	\label{eq:T}
	\DDy{T} &=& \frac{F}{\rho\kappa}\\
	\label{eq:F}
	\DDy{F} &=& C_{p} \dot{m} \left( \DDy{T} - 
	\frac{T}{y}\nabla_{s}\right) + \varepsilon.
    \end{eqnarray}
\end{mathletters}
Here $\nabla_{s}\equiv (\partial\ln T/\partial\ln p)_{s}$ is the
adiabatic gradient and $\lambda_{3\alpha} =
\varepsilon_{3\alpha}/Q_{3\alpha}$, where $Q_{3\alpha}=7.274\usp\MeV$
is the energy released per formed carbon.  Equations
(\ref{eq:He})--(\ref{eq:F}) depend on both the surface gravity $g$
(through $\rho(p=gy,T)$) and the local accretion rate $\dot{m}$.  We
fix $g = GM/R^{2} (1-2GM/Rc^{2})^{-1/2}$ to the value
$2.3\ee{14}\usp\cm\usp\second^{-2}$ appropriate for a neutron star of
mass $1.4\usp\Msun$ and radius $10\usp\kilo\meter$.

Because the accretion rate varies, we first fix the temperature of the
deep crust by solving equations (\ref{eq:He})--(\ref{eq:F}) at the mean
accretion rate $\dot{m}=2.25\ee{4}\usp\GramPerSc\usp\second^{-1}$, which
roughly corresponds to the persistent mean accretion rate of the source
if uniformly deposited over the surface.  To fix the composition of the
atmosphere and the bottom of the He-rich layer, we compute the burning
of He to C via the 3$\alpha$ reaction.  At low accretion rates, the
$\mathrm{^{12}C(\alpha,\gamma)^{16}O}$ reaction rate is negligible
compared to the 3$\alpha$ rate \citep{bildsten95:_propag}, and we
therefore just track the abundances of $\mathrm{^{4}He}$ and
$\mathrm{^{12}C}$.  For $\varepsilon_{3\alpha}$, the heating rate from
the $3\alpha$ reaction, we use the analytical fit of
\citet{fushiki87:_s}.

Boundary conditions are specified at both ends of the computational
domain.  The thermal structure of the atmosphere is insensitive to the
photosphere temperature (it is, roughly speaking, a radiative-zero
solution) for $y\gg K^{-1}$, $K$ being the opacity; for definiteness, we
set $T=(GM\dot{M}/2R\sigma_\mathrm{R})^{1/4}$ at $y =
10^{2.5}\usp\GramPerSc$, where $\sigma_\mathrm{R}$ is the
Stefan-Boltzmann constant.  The choice of boundary condition at the
other end of the computational domain (set to $y =
10^{14}\usp\GramPerSc$) is more problematic.  The high thermal
conductivity of the degenerate, relativistic electrons ensures that the
ocean is well-coupled to the thermal state of the inner crust and core.
Previous studies \citep[e.g.,][]{taam78:_nuclear,brown98} simply took
the flux to be zero at the inner boundary.  For an accreting neutron
star, continual compression of the deep crust forces a series of
electron captures, neutron emissions, and pycnonuclear reactions
\citep{bisnovatyi-kogan79:_noneq_x,blaes90:_slowl,haensel90a}; in
steady-state, most of the heat released from these reactions flows into
the core but a small amount ($\lesssim 0.1\usp\MeV$ per accreted nucleon
for $\dot{M}\gtrsim 0.1\dot{M}_\mathrm{Edd}$) flows outward
\citep{brown:nuclear}.  The thermal timescale in the inner crust is of
order years \citep{brown98:transients} and so the temperature there
depends mostly on the time-averaged accretion rate.  We therefore adjust
the temperature at $\log(y)=14.0$ so that the flux from the deep crust
is $\approx 0.1 \NA\dot{m} \usp\MeV$.

With these approximations, we numerically integrate the equations
(\ref{e:cont})--(\ref{e:fluxa}) using a stiff ODE solver.  We
integrate from each boundary and iteratively adjust the flux at the
upper and lower boundaries so that the temperature and flux are
continuous at a fitting point.  The result of this integration, over
the range of $y$ relevant for ignition, is shown in
Figure~\ref{fig:structure}.  We show the temperature (\emph{solid
line}; left-side vertical axis) and flux in units of MeV per accreted
nucleon (\emph{dotted line}; right-side vertical axis).  The arrow is
to guide the eye and mark the value of $T(y=10^{14}\usp\GramPerSc)$. 
The top axis shows the time required to accrete a given column at
$\dot{m}=2.3\ee{4}\usp\GramPerSc\usp\second^{-1}$.  As we shall
explain momentarily, it is plausible that the carbon and iron form a
mixed layer; we therefore change the mixture from
$\mathrm{^4He/^{12}C}$ to $\mathrm{^{12}C/^{56}Fe}$
($X_\mathrm{C}=0.3$) at $y=10^{10}\usp\GramPerSc$.  The kink in the
temperature at that point is because the conductivity decreases when
we reset the composition.  We chose $X_\mathrm{C}=0.3$ as a minimal
case for ignition.  A larger $X_\mathrm{C}$ would increase the ocean
thermal conductivity and, for a fixed flux of $0.1\usp\MeV\NA\dot{m}$,
lower the ocean temperature.  This increases both the recurrence time
and the total burst energetics (although, as we shall explain in the
next section, a larger $X_\mathrm{C}$ will not increase the X-ray
fluence substantially because the additional energy is lost to
neutrinos).  A smaller $X_\mathrm{C}$ may not ignite unstably
\citep{cumming.bildsten:carbon}.  With a judicious use of hindsight,
we set the base of the C/Fe layer at $y=10^{13}\usp\GramPerSc$
(roughly where the $\mathrm{^{12}C}$ will ignite); at greater
column the composition is pure Fe.  The point where the pure Fe layer
begins is marked in Figure~\ref{fig:structure} by the large black dot.

As explained in \citet{brown98}, thermal conduction is more efficient at
cooling the ocean than neutrino emission at these temperatures
\citep{brown98}.  The condition for ignition is crudely expressed as
$d\varepsilon_\mathrm{C}/dt > d\varepsilon_\mathrm{th}/dt$, where
$\varepsilon_\mathrm{th} = \rho\kappa T/y^{2}$ is an approximation to
$\rho^{-1} \mathbf{\nabla\cdot F}$.  For the heating
$\varepsilon_\mathrm{C}$, we use the rate of \citet{caughlan88:_therm}
and incorporate strong screening using the formalism of
\citet{ogata93:_therm}.  In Figure~\ref{fig:structure}, we show the
points where ignition occurs (\emph{heavy dashed curve}).  Note that the
strong temperature sensitivity of the $\mathrm{^{12}C}+\mathrm{^{12}C}$
reaction means that the location of this curve is relatively insensitive
to the precise method of defining ignition.  As is evident from the
figure, the temperature in the carbon-rich layer is too cold, by about a
factor of 2--3, for ignition to occur if the base of the carbon-rich
layer is at $y= 10^{11}\usp\GramPerSc$.  The presence of heavy elements
(e.g., $\mathrm{^{104}Ru}$) in the ocean of the other super burst
sources makes the carbon-containing layer hotter, for a given flux from
the crust and core, and allows for a smaller ignition column
\citep{cumming.bildsten:carbon}.

In this calculation, the mass of carbon at
the depth where ignition occurs is $\approx 4\pi R^2\times 0.3\times
10^{13}\usp\GramPerSc\approx 3.8\ee{25}\usp\gram$.  This is about a
factor of 20 greater than that indicated by our estimate of the burst
energetics.  For a lesser $X_\mathrm{C}$, the conductivity is lower
and the temperature at which ignition occurs increases; as pointed out
by \citet{cumming.bildsten:carbon}, however, the carbon burning will
become stable at low accretion rates and small carbon abundances. 
Their estimate implies that for the conditions we assume here,
$X_\mathrm{C}\gtrsim 0.5$ is required for unstable ignition. 
Consistent with this estimate, we find that for this calculation the
lifetime, $-X_\mathrm{C}/\dot{X}_\mathrm{C}$, is slightly less than
the accretion timescale $y/\dot{m}$ at the ignition point.  For this
reason we do not consider smaller mass fractions of carbon.

The temperature at the base of the carbon layer is mostly determined by
the flux flowing from deeper in the star.  The reason is that the heat
flowing from the crust ($\sim 0.1\usp\MeV$ per accreted nucleon) is
larger than that generated by compressional heating ($\sim
C_{p}T\nabla_{s}/\NA\sim 1\usp\keV$ per accreted nucleon), and the high
conductivity of the degenerate plasma enforces a very shallow
temperature gradient.  At the time of the super burst, the accretion
rate was lower than the mean, and the deep ocean was slightly colder
than in this calculation, making ignition even more difficult.

As mentioned earlier, locally high accretion rates onto a small
portion of the stellar surface may be an explanation for the lack of
bursts during the high-luminosity state of this source.  At higher
local accretion rates, the ignition column decreases.  To heat a
confined carbon column to ignition at $y=10^{11}\usp\GramPerSc$
requires local accretion rates in excess of 5 times the local
Eddington rate \citep{brown98}.  It is implausible, however, that the
accreted matter would fail to spread over the surface prior to
reaching this column.  While strong magnetic fields ($\sim
10^{12}\usp\gauss$) could possibly confine such a large column
\citep{brown98,litwin.brown.ea:ballooning}, there is no evidence, such
as pulsations in the persistent emission, for such strong organized
fields in this source.

What about heating of the carbon layer during a type I X-ray burst? 
The timescale to conduct the heat from the base of the He-rich layer
to the surface is $\sim 10\usp\second$ (the decay timescale of the
burst).  Unlike in the degenerate ocean, the dominant opacity is
Thompson scattering, for which the equivalent radiative conductivity
is independent of depth.  As a result, the inward propagation of heat
is inefficient (as can be checked by a calculation similar to that
described in equation~(\ref{eq:diffusion}) and the heating of the
carbon layer is too small to force an ignition.

The fact that this source cyclically produces $\mathrm{^{12}C}$ (in
the high-luminosity) state and iron-peak elements (through unsteady He
burning in the low-luminosity state) has some interesting
consequences.  During a He burst, the burning produces mostly
$\mathrm{^{56}Fe}$.  As the iron cools, it becomes denser than the
underlying carbon, because the iron has fewer electrons per nucleon
than helium.  The base of the unstable He burning layer is therefore
susceptible to first a secular salt-finger and then a Rayleigh-Taylor
instability.  For example, at an interface pressure of
$2\ee{23}\usp\dyne\usp\cm^{-2}$ and a $\mathrm{^{12}C}$ temperature of
$2.5\ee{8}\usp\K$, the iron layer is less dense than the carbon layer
only when its temperature is $> 7.6\ee{8}\usp\K$.  

Terrestrial Rayleigh-Taylor experiments and numerical simulations
\citep[and references therein]{youngs:RT} suggest that the time $t$ for
the interface (between fluids of density $\rho$ and $\rho'$) to traverse
a distance $s$ is given by $t = [s/(\alpha \mathcal{A} g)]^{1/2}$, where
$\mathcal{A} = (\rho-\rho')/(\rho+\rho')$ is the Atwood number and
$\alpha\approx 0.03$ is uncertain by a factor of 2.  Taking $s$ to be a
scale height, we find that for this case $t \sim 10^{-4}\usp\second$.
Thermal diffusion is ineffective on this timescale; the characteristic
distance over which heat can diffuse on this timescale is less than
$1\usp\cm$, which is much smaller than the pressure scale height ($\sim
1000\usp\cm$).  The scaling of distance with $t^{2}$ comes about because
of mergers between falling spikes into larger and larger drops; in the
case of a stratified medium, this effect is counteracted to some extent
by compression of the falling spikes.  As the iron ``plume'' descends,
the ambient medium (carbon) is entrained, reducing the buoyancy contrast
and eventually halting its descent \citep{townsend:entrainment}.

Because the mass of Fe available after a He burst is much less than
the mass of the carbon-rich layer, it is unlikely that the sinking Fe
will fall to the base of the carbon-rich layer.  Whether a descending
plume would be able to spark an ignition of the carbon layer at $y\ll
10^{11}\usp\GramPerSc$, and whether a flame could propagate without
quenching at those lower densities, is unknown.  In any case, it is
likely that at least some of the iron is mixed into the carbon layer
and advected deeper into the star by continual accretion, so that
ignition will occur in a mixed C/Fe layer, as we have assumed.  
Detailed calculations are outside the scope of this paper; for now we 
leave this as an interesting possibility.

\subsection{Evolution of the Burst}\label{subsec:thermal-decay}

As noted above, the ignition of $\mathrm{^{12}C}$, for conditions
relevant to this source, requires either a large accumulated column of
carbon-rich material, or a trigger, perhaps associated with an earlier
He flash.  In this subsection, we explore the evolution of an unstable
ignition at $y=10^{13}\usp\GramPerSc$ of a C/Fe layer
($X_\mathrm{C}=0.3$).  Unlike for normal (10--20~s duration) type I
X-ray bursts, the energetics of this super burst are much greater than
that inferred from the X-ray lightcurve.  There are two reasons for
this.  The first is that neutrino cooling is more efficient than
thermal conduction at the peak temperature ($>5\ee{9}\usp\K$) reached
in the burst.  The second reason is that thermal conduction into
deeper layers is efficient and competes with the outward thermal
transport.  Because of the great depth of ignition, the thermal time
is of order $10^5\usp\second$ and is much longer than the duration of
the observation.  We now consider each of these cooling mechanisms in
turn.

Once the instability begins, the heating timescale
$C_{p}T/\varepsilon_\mathrm{C}$ is, by definition, shorter than both the
thermal conduction and neutrino cooling timescales.  At first the
nuclear burning timescale is comparable to the thermal conduction time
($\gtrsim 10^4\usp\second$; \citealt{taam78:_nuclear}), but it
accelerates with increasing temperature.  We calculate the peak
temperature from the equation $\int_{T_{i}}^{T_{p}} C_{p}\,dT =
X_\mathrm{C} E_\mathrm{nuc}$, where $E_\mathrm{nuc} \approx
10^{18}\usp\erg\usp\gram^{-1}$ is the energy per gram released in
burning $\mathrm{^{12}C}$ to Fe-peak elements.  For a pure carbon layer,
the temperature would reach $T_p \approx 1.5\ee{10}\usp\K$ if all the
carbon burned to iron.  Note that even at this extreme temperature, the
electrons are still (mildly) degenerate; the Fermi energy is $\approx
5\usp\MeV$.  In addition, the radiation pressure, at this temperature,
is only $1.3\ee{26}\usp\dyne\usp\cm^{-2}$, about a tenth of the total.
Unlike in He burning, the peak temperature is therefore not limited by
the expansion of the burning layer if the base of that layer is sufficiently
deep.  Because of the high density, photodisintegration reactions are
also unimportant until temperatures in excess of $10^{10}\usp\K$ are
reached, as can be checked by a Saha-type equation.  For the abundance
($X_\mathrm{C}=0.3$) used in Figure~\ref{fig:structure}, $T_p =
9.1\ee{9}\usp\K$.

Although neutrino cooling is unimportant for stabilizing the ignition of
carbon on a neutron star \citep{brown98}, it is more efficient than
thermal conduction when the burning layer is at its peak
temperature\footnote{This point was also made by
\citet{cumming.bildsten:carbon} and we thank L. Bildsten for stressing
this to us.}.  In addition, thermal conduction to the surface is more
efficient at lower column and competes more effectively with neutrino
cooling.  For a very deep layer, however, the situation is different.
The thermal conduction timescale is longer and the peak temperature
reached is much higher.  To investigate the cooling, we
constructed a one-zone approximation of the layer
following the burning,
\begin{equation}
    C_{p}\ddt{T} = -\frac{\rho\kappa T}{y^{2}} - \epsilon_{\nu}.
    \label{eq:one-zone-neutrino}
\end{equation}
We set the initial temperature at $T_{p}(X_\mathrm{C})$ but switched
the composition to Fe to simulate the ashes.  At a temperature $\sim
10^{10}\usp\K$, the dominant neutrino-producing process is from
electron-positron pairs.  As a result, the neutrino emissivity
\emph{decreases} as density increases because of the rising electron
Fermi energy (see Fig.~\ref{fig:enu}), and it is likely that a strong
temperature gradient develops, so this one-zone calculation
underestimates somewhat the neutrino flux.  Our calculation is also
limited in that it completely neglects convection.

In Figure~\ref{fig:compare-fluence}, we show the neutrino (\emph{solid
line}) and conductive (\emph{dotted line}) fluence, as a function of
time, computed with equation~(\ref{eq:one-zone-neutrino}) for a
fraction of carbon $X_\mathrm{C}=0.3$.  For $X_\mathrm{C} > 0.3$, the
total conductive fluence of the burst only increases slightly; any
further heating is carried off by neutrinos.  For this calculation,
the total radiated energy is $\approx 1.5\ee{43}\usp\erg$ for a
neutron star of $10\usp\kilo\meter$ radius---about a factor of 10
greater than that observed, although for this calculation the total
energy conducted to the surface after $10^{4}\usp\second$ is only
$1.6\ee{42}\usp\erg$, consistent with that measured.  The evolution of
the luminosity is shown in Figure~\ref{fig:luminosity}.  For such a
deep layer of carbon, the thermal evolution time is quite long; the
burst flux still exceeds that from accretion for several \emph{days},
in agreement with calculations by \citet{cumming.bildsten:carbon} for
super bursts from hydrogen accreting sources.

While a one-zone approximation is tolerable at lower columns, it
becomes suspect where the electrons are degenerate, since the
conductivity increases with density.  The electron-thermal
conductivity is
\begin{equation}
    \kappa = \frac{\pi^{2}}{3}\frac{n_{e}\kB^{2}T}{m_{e}^{\star}}\tau
    \label{eq:K}
\end{equation}
where $\tau = p_\mathrm{F}^{2}v_\mathrm{F}/(4\pi
Z^{2}e^{4}n_{N})\Lambda_{ei}$ is the electron-ion relaxation time.  Here
$p_\mathrm{F}$ and $v_\mathrm{F}$ are the electron momentum and velocity
evaluated at the Fermi energy $\EF$, $n_{e}$ and $n_{N}$ are the
densities of electrons and nuclei, respectively, $m_{e}^{\star} =
\EF/c^{2}$ is the relativistic electron mass and $\Lambda_{ei}$ is the
Coulomb logarithm.  Where relativistic, degenerate electrons dominate
the pressure, the conductivity scales as $\kappa \propto T p^{1/4}$
(neglecting variation in $\Lambda_{ei}$).  Because the thermal
conductivity increases with depth (pressure), heat deposited from
unstable carbon burning can readily flow inwards.

To illustrate the inward flow of heat, we construct the simplest thermal
transport problem by neglecting the variation of $\Lambda_{ei}$, taking
the pressure to be solely that from the degenerate electrons, and
writing the specific heat as $\propto \kB T/\EF$ (since the electrons
are more numerous than the ions, this is an adequate approximation for
most of the cooling).  With these simplifications, the thermal
diffusion equation in non-dimensional form is
\begin{equation}
    \frac{u}{x^{1/4}}\partial_{t}u - \partial_{x}(ux\partial_{x}u) = 0
    \label{eq:diffusion}
\end{equation}
over the domain $1 \le x \le x_R$.  Here $x = y/y_{\circ}$, $u =
T/T(y=y_{\circ})$, and the time is in units of
$C_{p}y^{2}/(\rho\kappa)|_{y=y_{\circ}}$.  The flux in these units is
$ux\partial_{x}u$.  The background steady-state solution is one of
constant flux,
\begin{equation}
    u_\mathrm{steady}(x,t=0) = \left[ 1 + \left(2\delta u + \delta u^{2}
    \right) \frac{\ln x}{\ln x_{R}} \right]^{1/2},
    \label{eq:uinit}
\end{equation}
where $\delta u$ is the rise in temperature over the domain.  To solve
equation~(\ref{eq:diffusion}) numerically, we use the method of lines
(\textsf{PDECOL}; \citealt{madsen.sincovec:pdecol}), with a
fifth-order polynomial spatial decomposition over logarithmically
spaced collocation points and implicit integration in $t$.  For
simplicity, the temperatures at both ends of the computational domain
are fixed.

At time $t=0$ the temperature profile is $u_\mathrm{steady}$ plus a
smoothed top-hat profile representing the heating injected during
unstable burning.  In order to investigate the transport of heat
following a perturbation, we placed the boundaries at
$10^{9}\usp\GramPerSc$ and $10^{13}\usp\GramPerSc$ and chose the
top-hat perturbation to span
$1.5\ee{12}\usp\GramPerSc<y<3.5\ee{12}\usp\GramPerSc$, with an initial
temperature of $9.1\ee{9}\usp\K$.

Figure \ref{fig:fluence} shows the fluence, defined here as $\int_0^t
(|F| - F_{S})\,dt'$, $F_{S}$ being the steady-state flux, as a
function of time.  The solid curve denotes the fluence at the top of
the layer; the dotted curve denotes that at the bottom.  After
$5\ee{5}\usp\second$, only about 0.16 of the total fluence has been
radiated from the surface.  The inward-directed flux will raise the
temperature in the deep ocean and crust slightly and will therefore
increase the flux from the crust above the $\approx
0.1\usp\MeV\NA\dot{m}$ long-term value.  While all of the heat
deposited in the ocean will eventually be radiated from the surface,
only the fraction that is radiated immediately following the burst is
discernible in the X-ray lightcurve.  As in
Figure~\ref{fig:compare-fluence}, the evolution timescale is roughly
the thermal timescale at the base of the heated layer.

This simple calculation is only meant to be illustrative: it ignores
neutrino cooling and in particular it underestimates the thermal
transport to the surface.  A convective layer develops during the burst
\citep{taam78:_nuclear}; this convection may be enhanced by the stronger
neutrino emissivity at lower densities (Fig.~\ref{fig:enu}).  To
reproduce correctly the burst evolution and the lightcurve requires an
implicit evolution code, which is beyond the scope of this paper; for
now we note that the total fluence of the super burst is systematically
underestimated, by perhaps as much as an order of magnitude.

For both cooling calculations (eqs.~[\ref{eq:one-zone-neutrino}] and
[\ref{eq:diffusion}]), only a small fraction of the burst energy is
radiated away after $10^4\usp\second$.  \emph{As a result, the fluence
observed during the three hours following the burst is consistent with
the burning of a large mass of carbon at great depth.} The strongest
constraint on this scenario is the recurrence time: to accrete a
column of $10^{13}\usp\GramPerSc$ takes $13.3 /
(\dot{M}/3\ee{17}\usp\gram\usp\second^{-1})\usp\yr$.

\section{Discussion and Summary}

Thermonuclear bursts provide a unique probe of the physics of neutron
stars.  Their properties depend on the accretion rate and composition
of the accreted matter, the bulk properties of neutron stars, the
thermal state of the neutron star and the detailed nuclear burning
physics.  4U 1820-30 is an especially revealing system since its
extreme properties place strong constraints on the composition of the
accreted matter.  The observation of a super burst from 4U 1820-30
with $> 1,000$ times the energy release of a typical helium burst
provides us with a compelling look at processes occurring at much
greater depth in the neutron star ``ocean'' than we are normally
privileged to witness.  

We have investigated the ignition of a $\mathrm{^{12}C}$-rich layer,
presumably produced during the high-luminosity state when no type~I
X-ray bursts are observed.  When the mass of this layer is $\gtrsim
10^{26}\usp\gram$, a thermal runaway ensues.  Unlike the case discussed
by \citet{cumming.bildsten:carbon}, the lack of heavy rp-process ashes
makes the carbon layer cooler for this source and a larger mass of
carbon is required for ignition.  Once the runaway begins, the
temperature rises to $\sim 10^{10}\usp\K$.  The burning ends and the
layer rapidly cools by neutrino emission until the temperature is $\sim
5\ee{9}\usp\K$, at which point thermal conduction becomes more efficient
at cooling the layer.  Because of the great depth of the ignition, heat
readily flows inward, and the temperature evolves on a timescale $\sim
10^5\usp\second$.  As a result, the measured X-ray fluence is only a
small fraction of the total, so that this scenario is consistent with
the observation.  The recurrence time would provide a strong constraint
on the ignition column.  To accrete an ignition column of
$10^{13}\usp\GramPerSc$ takes $\sim 13 /
(\dot{M}/3\times10^{17}\usp\gram\usp\second^{-1})\usp\yr$.  If the
accretion rate were much higher than that inferred by the X-ray
luminosity, than the ocean would be hotter, and the ignition column
reduced.  Recurrence times of less than a year would require that
$\dot{M}>10^{18}\usp\GramPerSc$, however, or that the ignition be
prematurely triggered.

As noted by \citet{cumming.bildsten:carbon}, the heating from this burst
can quench type I X-ray bursts until the cooling luminosity is $\approx
0.01 L_\mathrm{accr}$ \citep{bildsten95:_propag}.  If the scenario we
outline is correct, then no type~I X-ray bursts should have occurred in
the week following the super burst.  Distinguishing the elevated flux
following the burst will be more difficult, given the variability of the
accretion flux.  The strongest constraint on this scenario would be if
another super burst were detected after a timescale much less than a
decade.  This would suggest that the carbon ignition is triggered.  The
fact that a He burst occurred immediately prior to the super burst is
intriguing; if the carbon ignition did occur at large densities, then
the thermal instability began several hours prior to the observed rise
of the burst.  It is possible that the flux from the carbon runaway
could ignite any accumulated He.

While the calculations presented here are suggestive, they are very
crude and could be substantially improved.  Evolutionary calculations,
with reaction networks, of the ignition and subsequent evolution of this
burst, coupled with similar calculations for the super bursts seen from
hydrogen accreting sources, can inform us about the relevant physics at
work in the deep ocean.  Continued X-ray monitoring of the source should
eventually give a better constraint on the recurrence time.  An
important quantity is the amount of carbon produced; for the scenario we
outline here to work, roughly half of the accreted matter must burn to
carbon so that the ignition is indeed unstable.

The X-ray spectrum during the burst reveals a broad emission line
indicative of Fe K$\alpha$ fluorescence as well as an absorption edge in
the 7--9~keV range.  The very high signal-to-noise spectra obtained
during the burst make the detection of these discrete features extremely
secure.  Previous authors have predicted that such features might be
produced by disk reflection (see for example Day \& Done 1991), but to
our knowledge this is the first confident detection.  Detailed
modeling of these features could provide a new probe of the accretion
disk in bursting systems.  Here we have only scratched the surface in
terms of investigating the discrete components.  An in-depth study and
interpretation of the line and edge will be presented in a future
publication.

\acknowledgements

It is a pleasure to thank Lars Bildsten, Alan Calder, Andrew Cumming,
Erik Kuulkers, Yuri Levin, Craig Markwardt, Richard Mushotzky, Bob
Rosner, Jean Swank, Frank Timmes and Jim Truran for many helpful
discussions.  We also thank Frank Timmes for providing the
interpolation routines and tables for the equation of state used in
the computations in this paper.  EFB acknowledges support from an
Enrico Fermi Fellowship.  This work is partially supported by the
Department of Energy under grant B341495 to the Center for
Astrophysical Thermonuclear Flashes at the University of Chicago.

\vfill\eject

\vfill\eject

\section{Figure Captions}

\figcaption[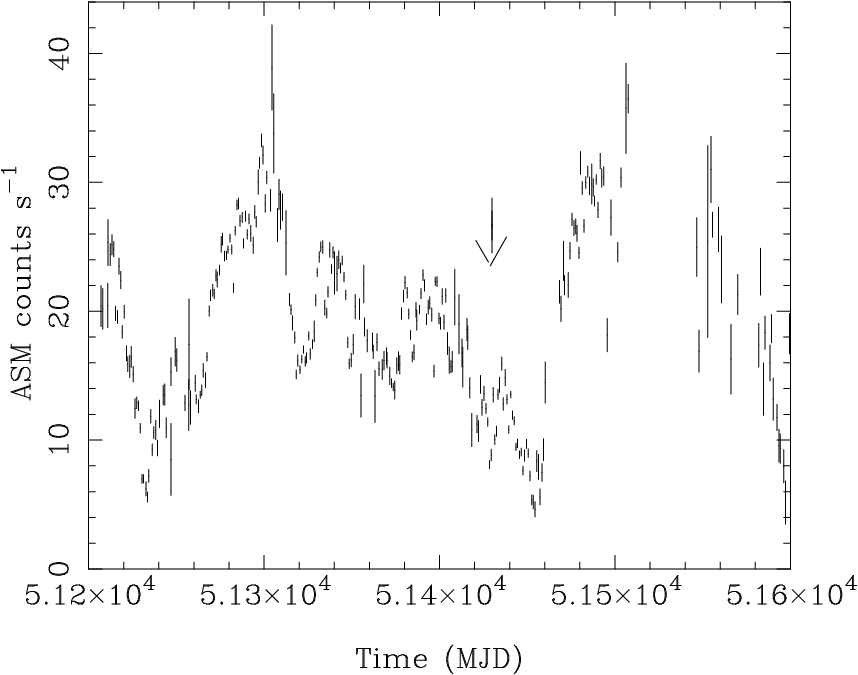]{RXTE/ASM lightcurve of 4U 1820-30 prior to 
and around the epoch of the super burst. A flux of 1 Crab is approximately 
75 ASM units.\label{fig1}} 

\vskip 10pt

\figcaption[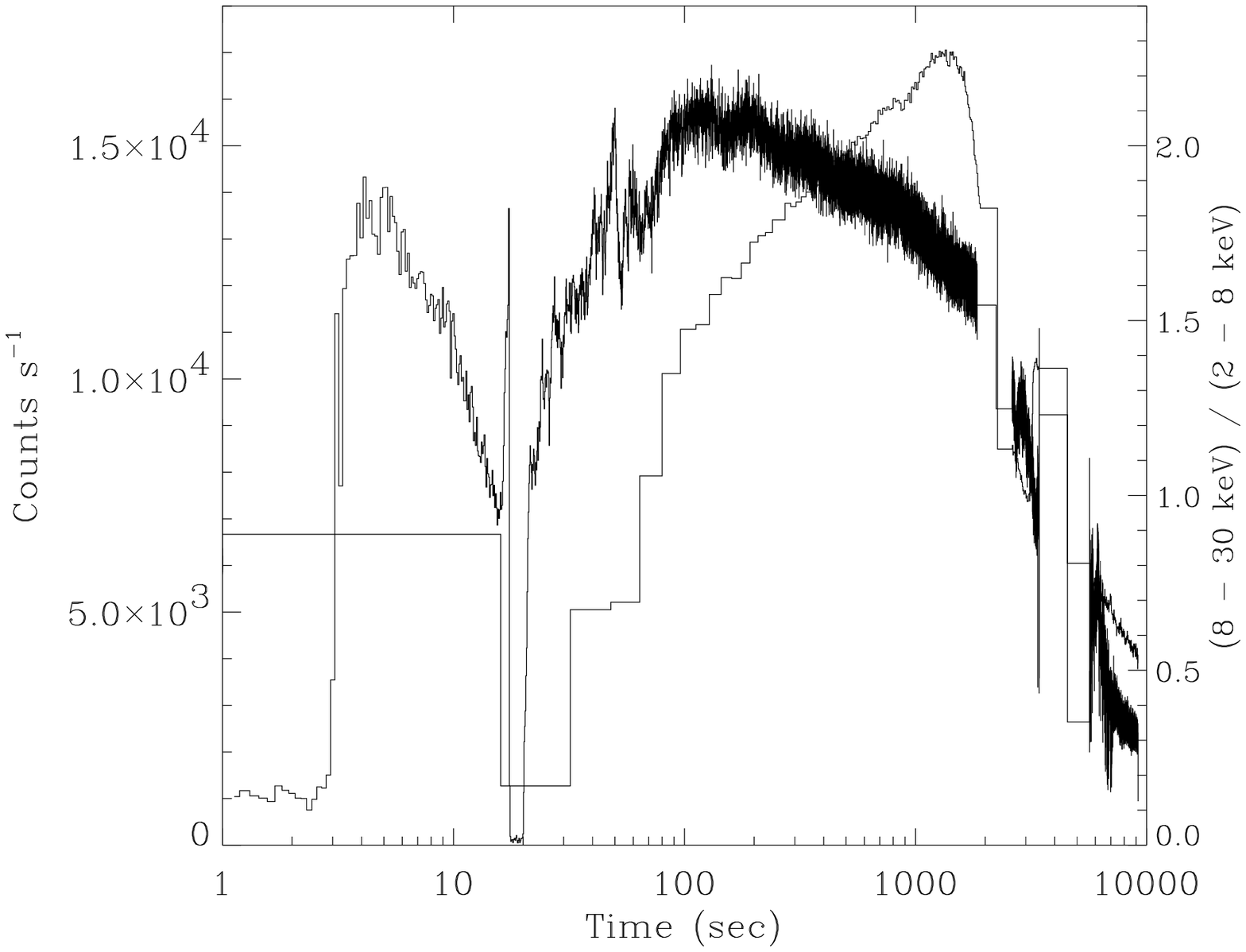]{Time history of the X-ray flux from 
4U 1820-30 during the super burst. The PCA lightcurve 
(2 - 90 keV) at 1/8 s resolution is the higher time resolution trace (left
axis). The lower time resolution curve is the (8 - 30)/(2 - 8) keV hardness 
ratio from Standard2 data with 16 s resolution (right axis). \label{fig2}}

\vskip 10pt

\figcaption[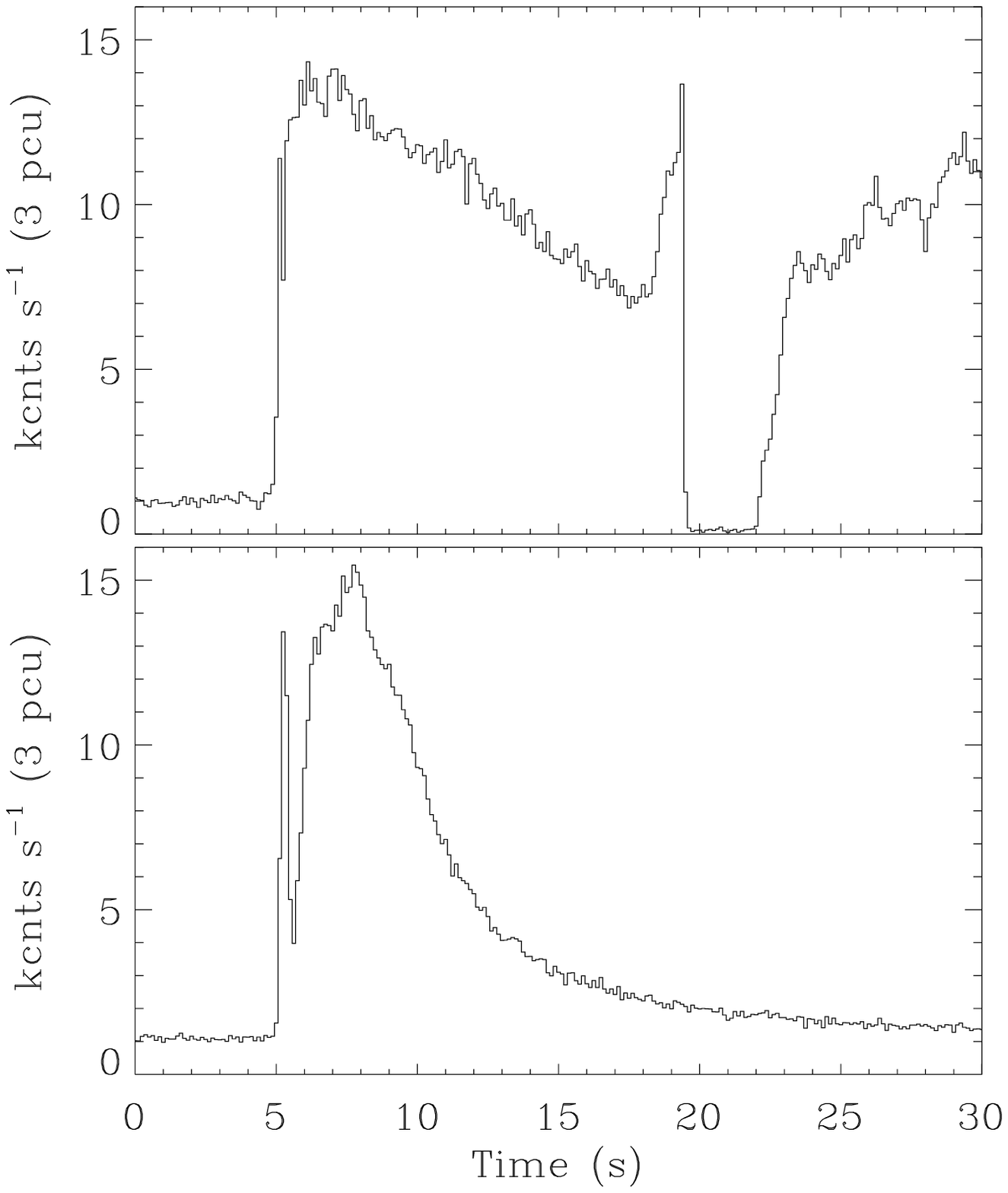]{Comparison of the helium flash which
preceded the super burst (top panel) and a helium flash which was observed on 
May 2, 1997 (bottom panel). Shown for both bursts are the 2 - 90 keV 
lightcurves from Standard1 data with 1/8 s time resolution. \label{fig3}}

\vskip 10pt

\figcaption[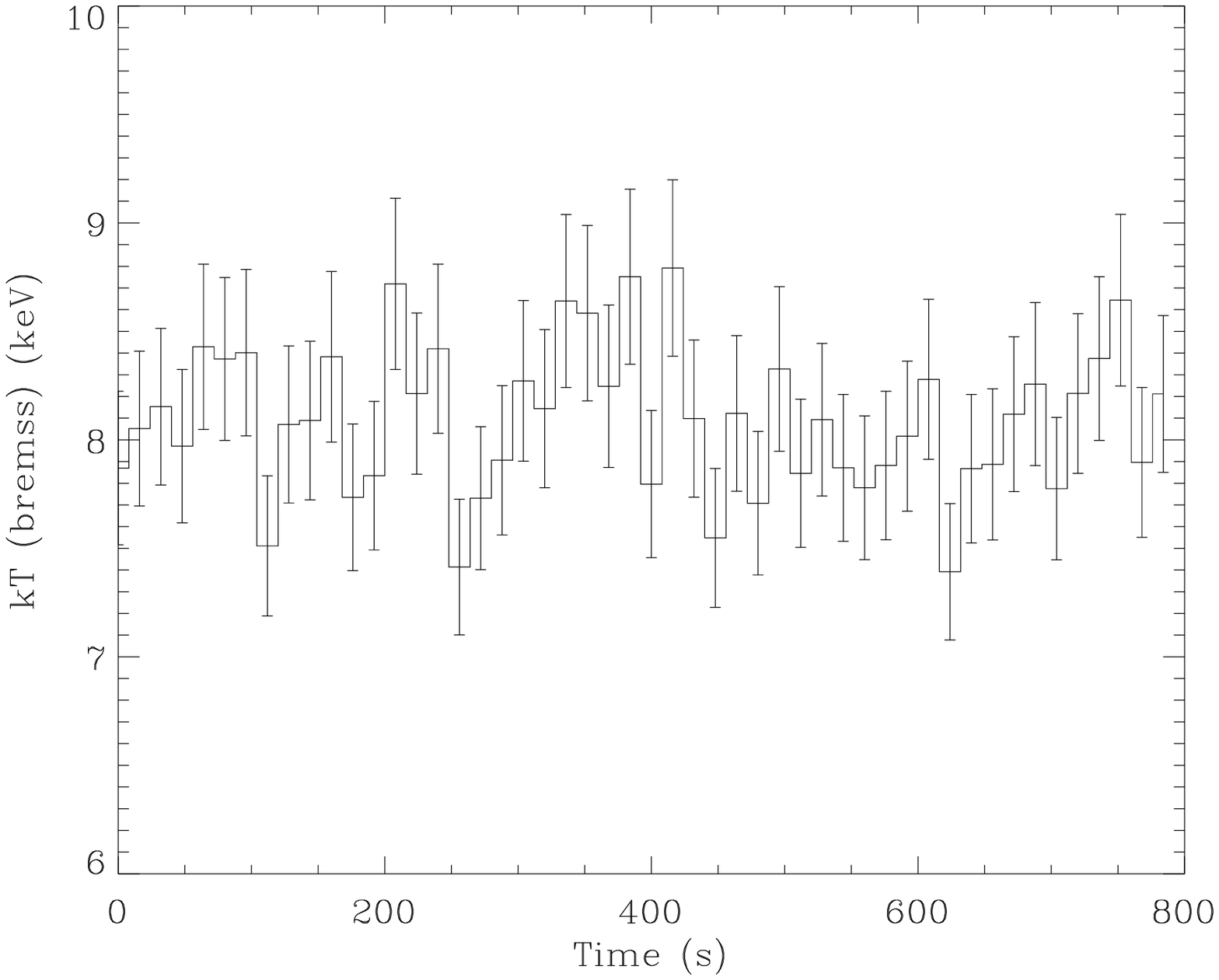]{Derived bremsstrahlung temperatures
(top panel) and 2 - 20 keV fluxes (bottom panel) for the persistent, accretion
driven flux prior to the super burst. \label{fig4}}

\vskip 10 pt

\figcaption[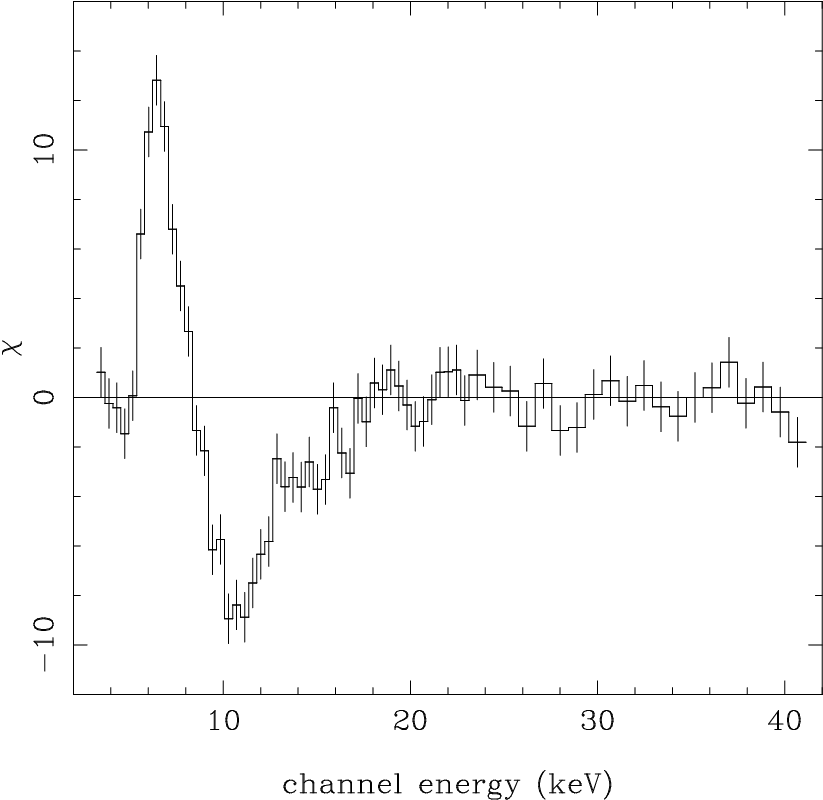]{Residuals to a spectral fit during the 
peak of the burst using a black body function with photoelectric absorption.
Shown are the residuals, data - model, in units of standard deviations. The
emission line feature just longward of 6 keV as well as the edge between 
8 - 10 keV are cleary evident. \label{fig5}}

\vskip 10 pt

\figcaption[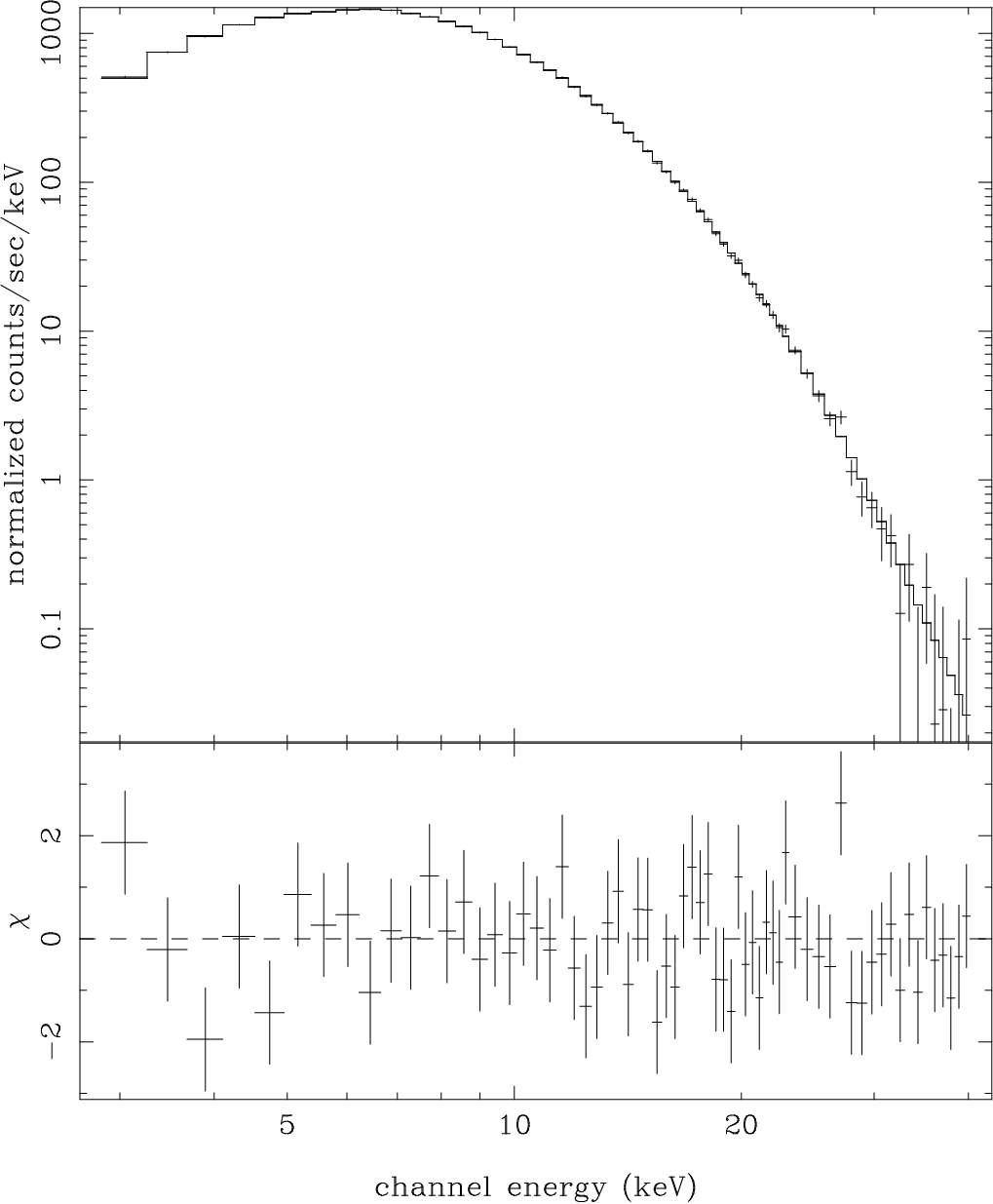]{Count rate spectrum and model (top 
panel) and residuals, data - model in units of standard deviations 
(bottom panel) for a spectral fit to an interval during the peak of the burst. 
The model includes a black body function, a gaussian emission line near 6 kev
and an absorption edge near 9 keV. Note that the error bars near the peak of
the spectrum are almost smaller than the width of the lines.  \label{fig6}}

\vskip 10 pt

\figcaption[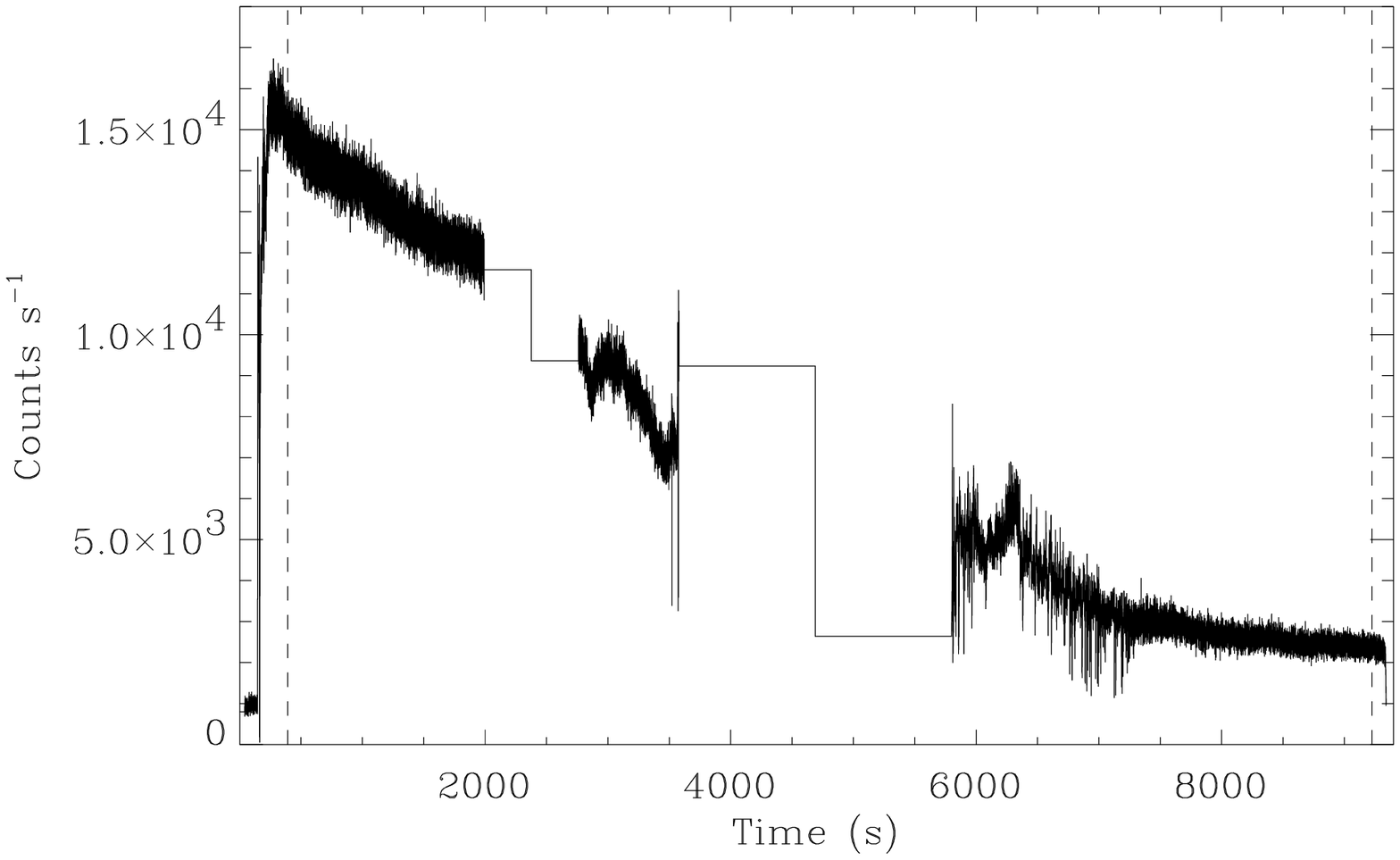]{Lightcurve of the burst from Standard1
data on a linear time axis. The dashed vertical lines denote the region in 
which we investigated the spectral evolution during the burst.  \label{fig7}}

\vskip 10 pt

\figcaption[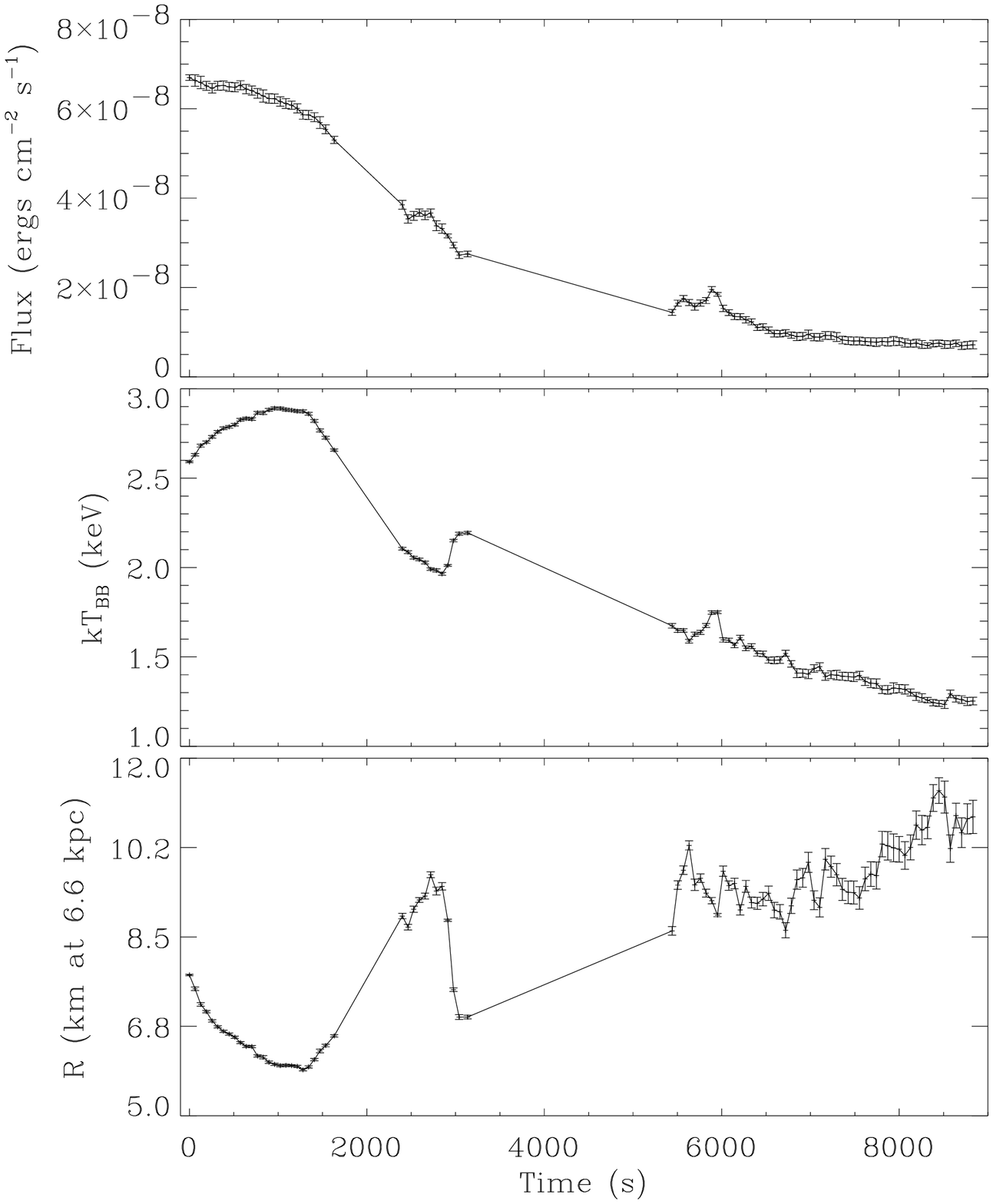]{Time evolution of the bolometric flux
(top panel), black body temperature (middle) and the inferred black body
radius (bottom) assuming a distance of 6.6 kpc. The time interval shown 
corresponds to the interval between the vertical dashed lines in figure 7.
The bolometric flux was derived by integrating the best-fit black body 
spectrum. \label{fig8}}

\vskip 10 pt

\figcaption[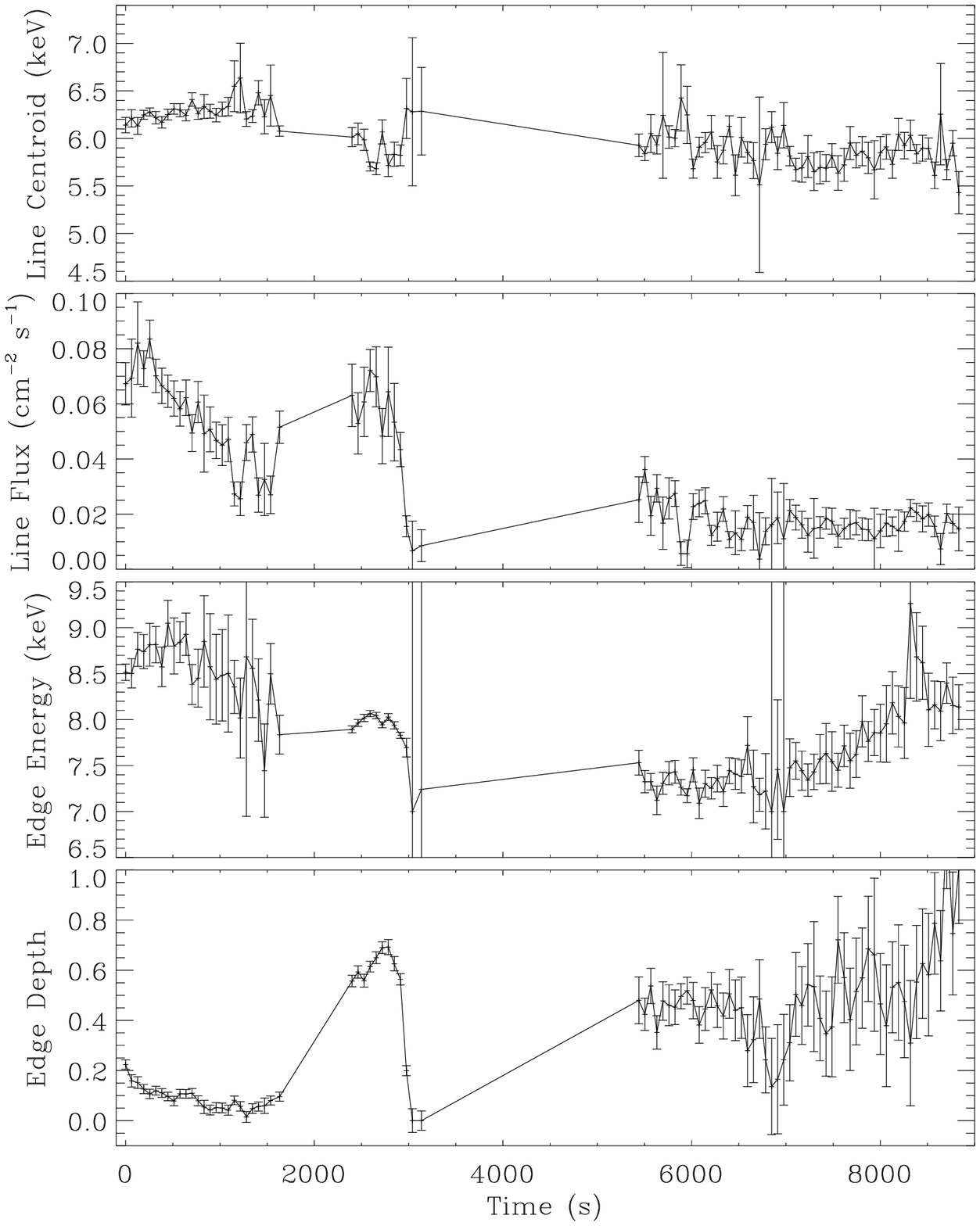]{Time evolution of the emission line and 
absorption edge components throughout the burst. Shown are the line centroid
energy, and line flux from the diskline model (top and 2nd from top) along with
the edge energy and depth (bottom and next to bottom) from the smedge model. 
\label{fig9}}

\vskip 10 pt

\figcaption[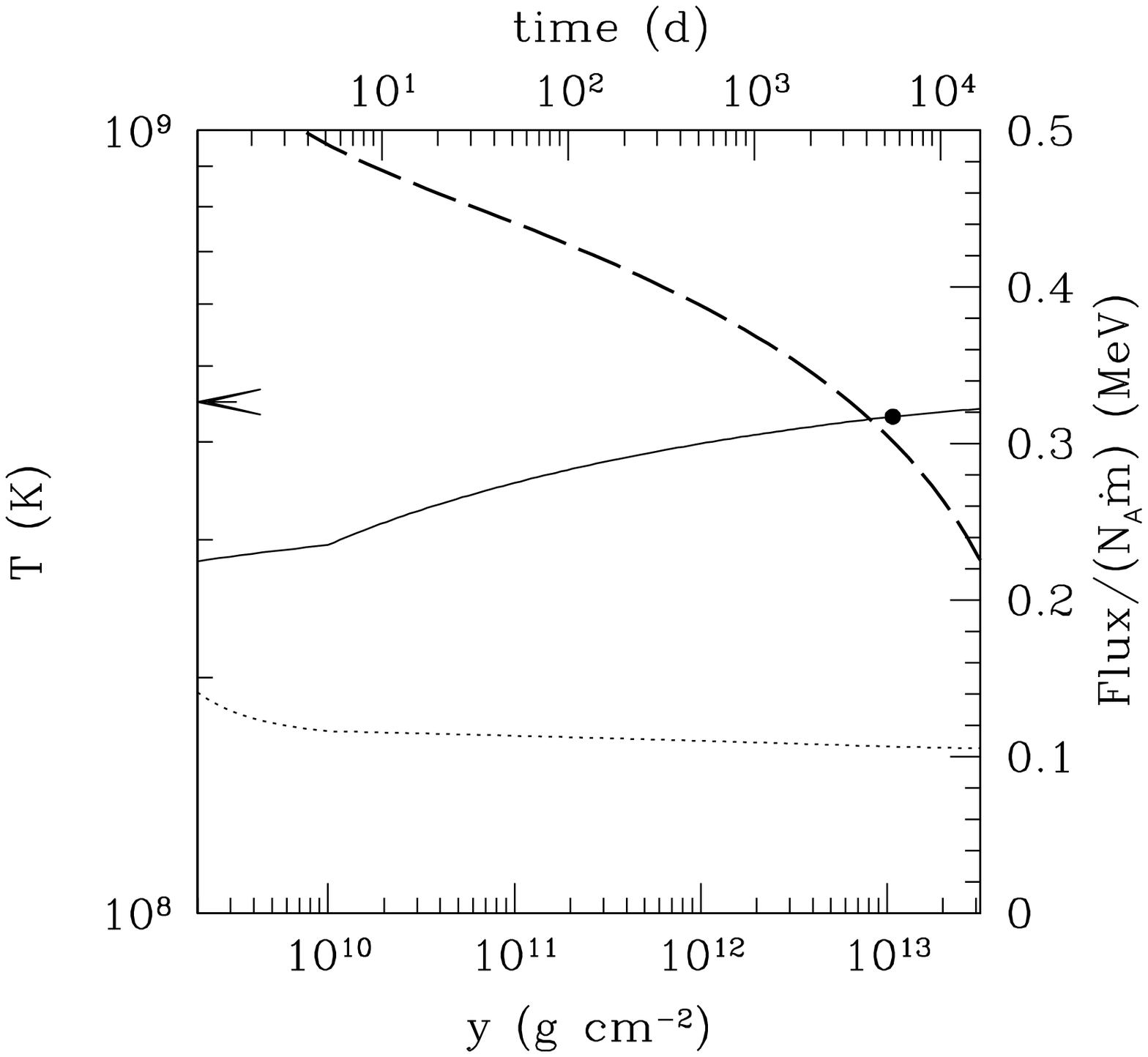]{The thermal structure of the other layers of a
neutron star accreting pure He at a local accretion rate
$\dot{m}=2.25\times 10^4\usp\GramPerSc\usp\second^{-1}$.  The top axis
indicates the time required for a column $y$ to be accreted at this
rate.  Plotted are the temperature (\emph{solid line}) and flux
(\emph{dotted line}) in units of MeV per accreted nucleon.  The arrow is
to guide the eye to the value of $T$ at $y=10^{14}\usp\GramPerSc$.  The
dots indicates the base of the C/Fe layer.  The condition for
instability of the reaction $\mathrm{^{12}C}+\mathrm{^{12}C}$ is
indicated by the heavy dashed curve.
\label{fig:structure}}

\vskip 10 pt

\figcaption[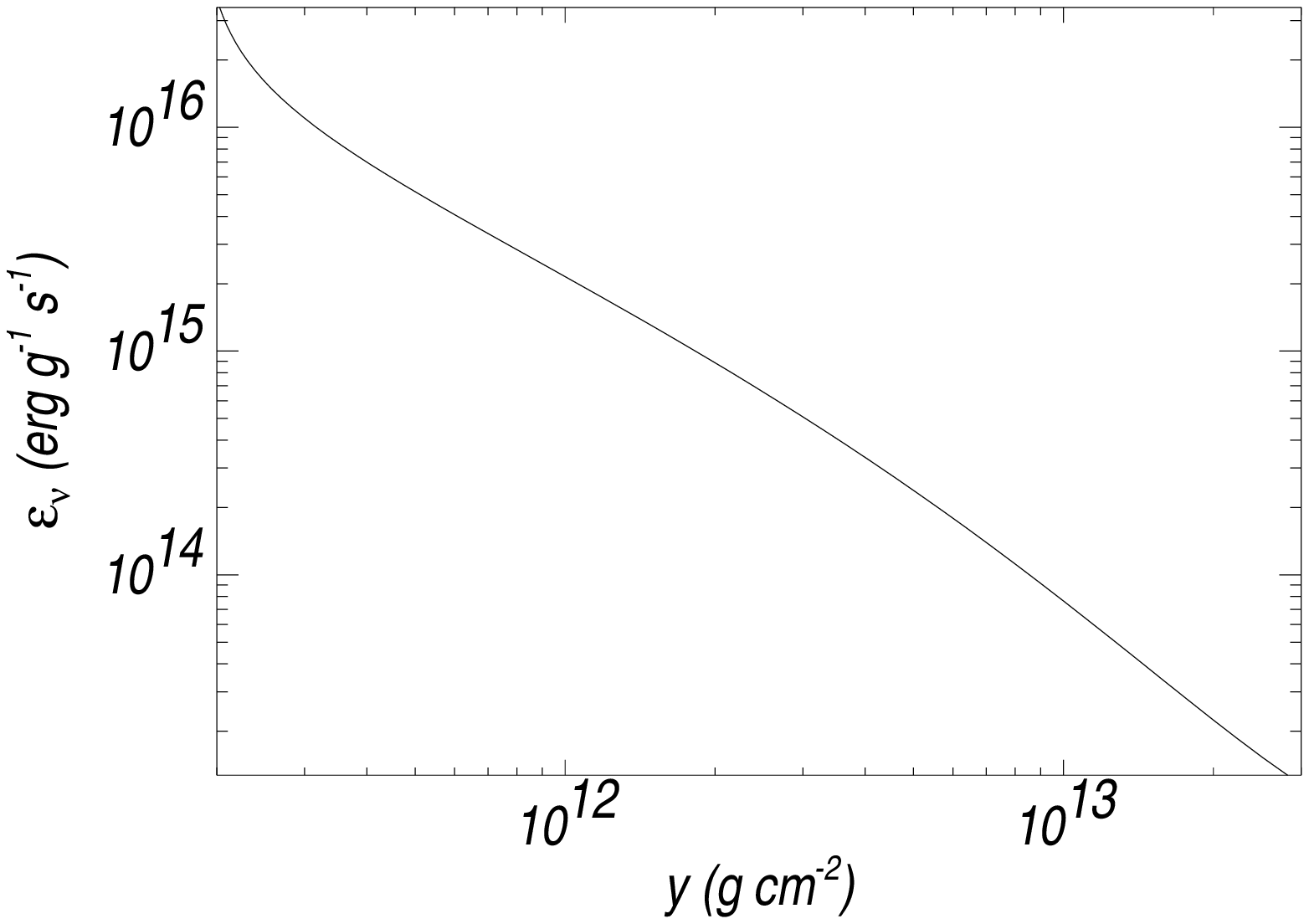]{The neutrino emissivity at a temperature of 
$10^{10}\usp\K$, as function of column for a surface gravity 
$2.34\ee{14}\usp\cm\usp\second^{-1}$.
\label{fig:enu}}

\vskip 10 pt

\figcaption[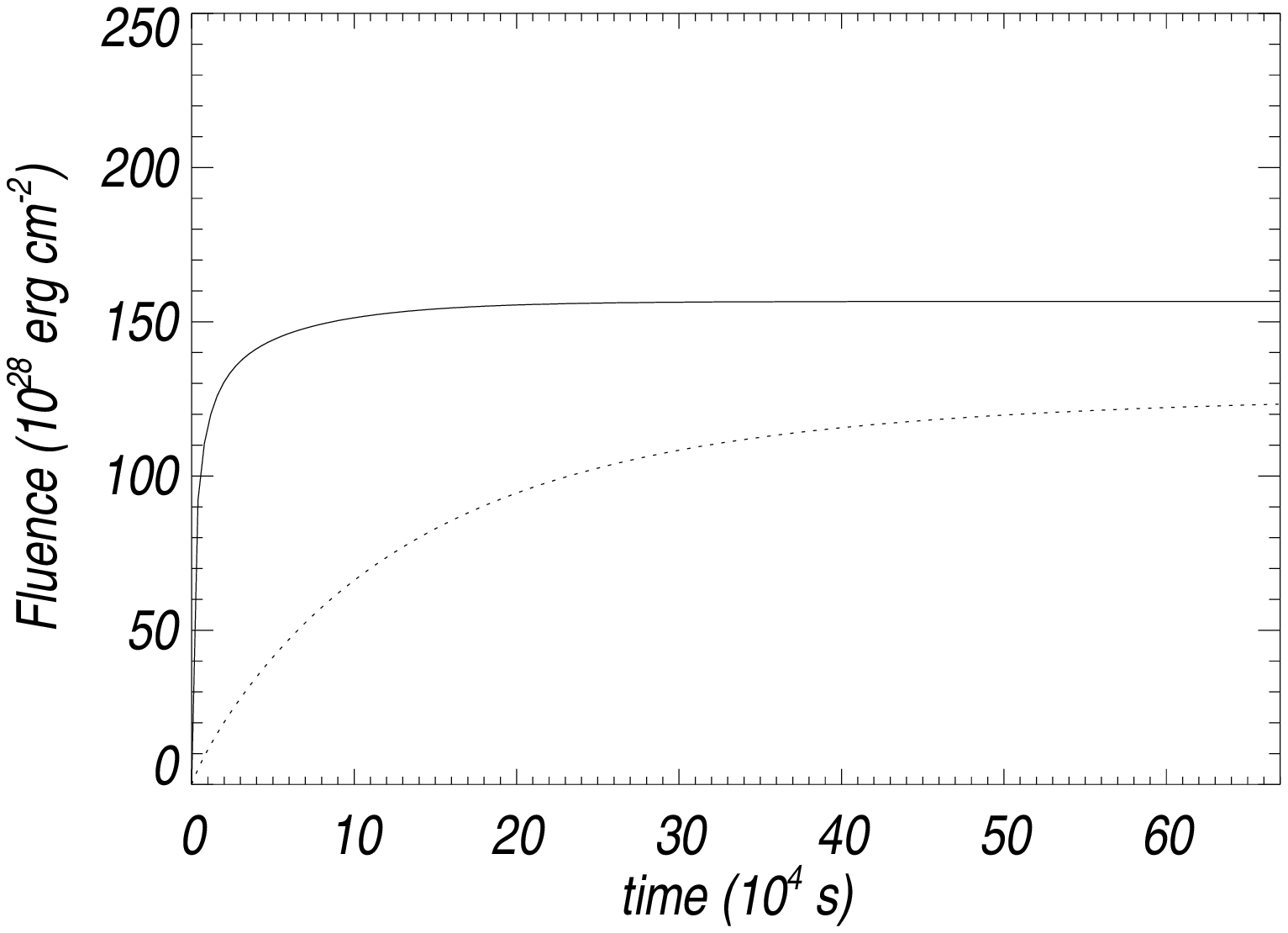]{The fluence, in units of
$10^{28}\usp\erg\usp\cm^{-2}$, as a function of time, in units of
$10^{4}\usp\second$, for the one-zone calculation
(eq.~[\protect\ref{eq:one-zone-neutrino}]).  Shown are both the neutrino
(\emph{solid line}) and conductive (\emph{dotted line}) fluences.  The
conductive flux evolves on the thermal diffusion timescale appropriate
for a depth of $y=10^{13}\usp\GramPerSc$.
\label{fig:compare-fluence}}

\vskip 10 pt

\figcaption[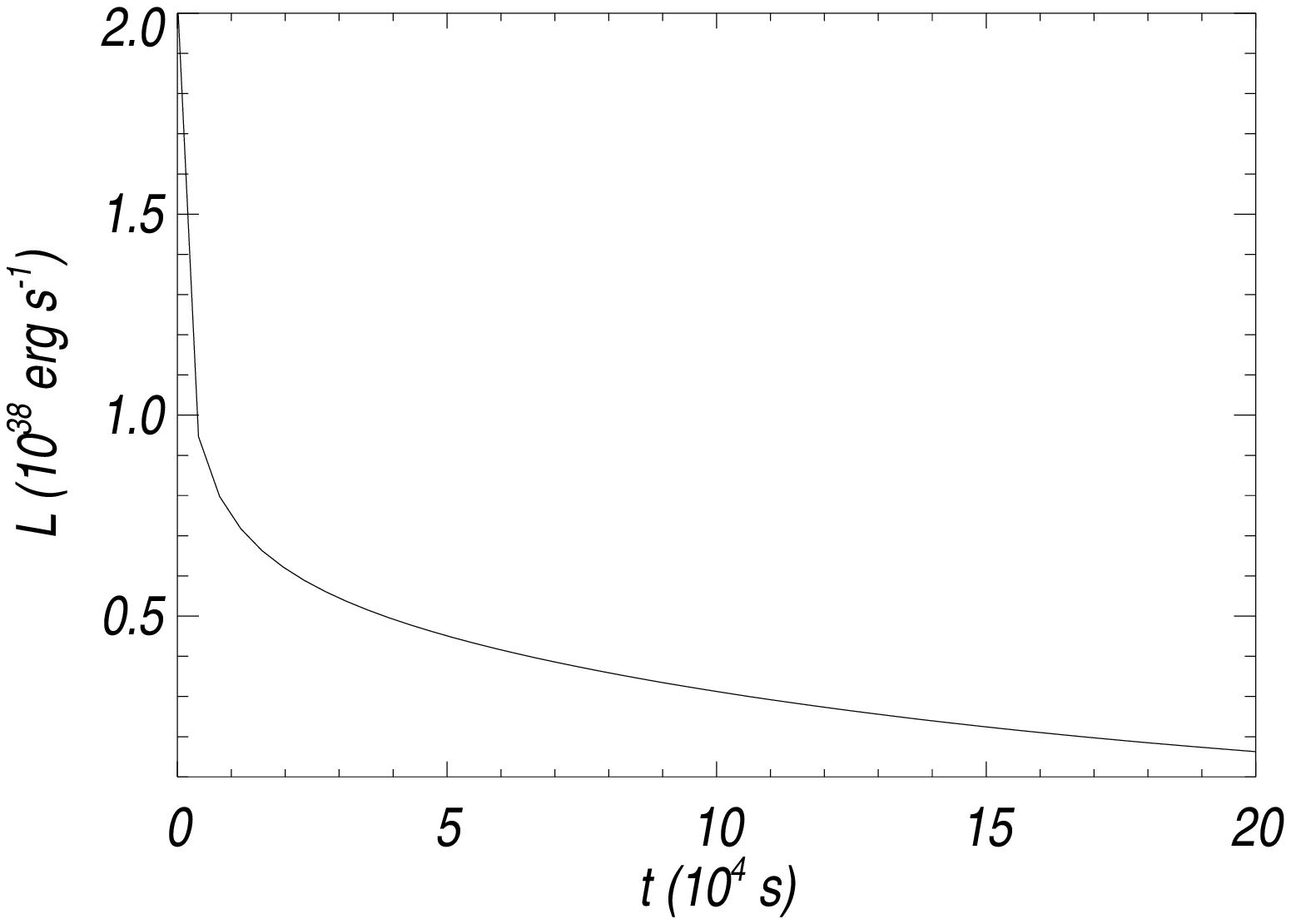]{The luminosity, in units of
$10^{38}\usp\erg\usp\second^{-1}$, as a function of time, in units of
$10^{4}\usp\second$, for the one-zone calculation
(eq.~[\protect\ref{eq:one-zone-neutrino}]).
\label{fig:luminosity}}

\vskip 10 pt

\figcaption[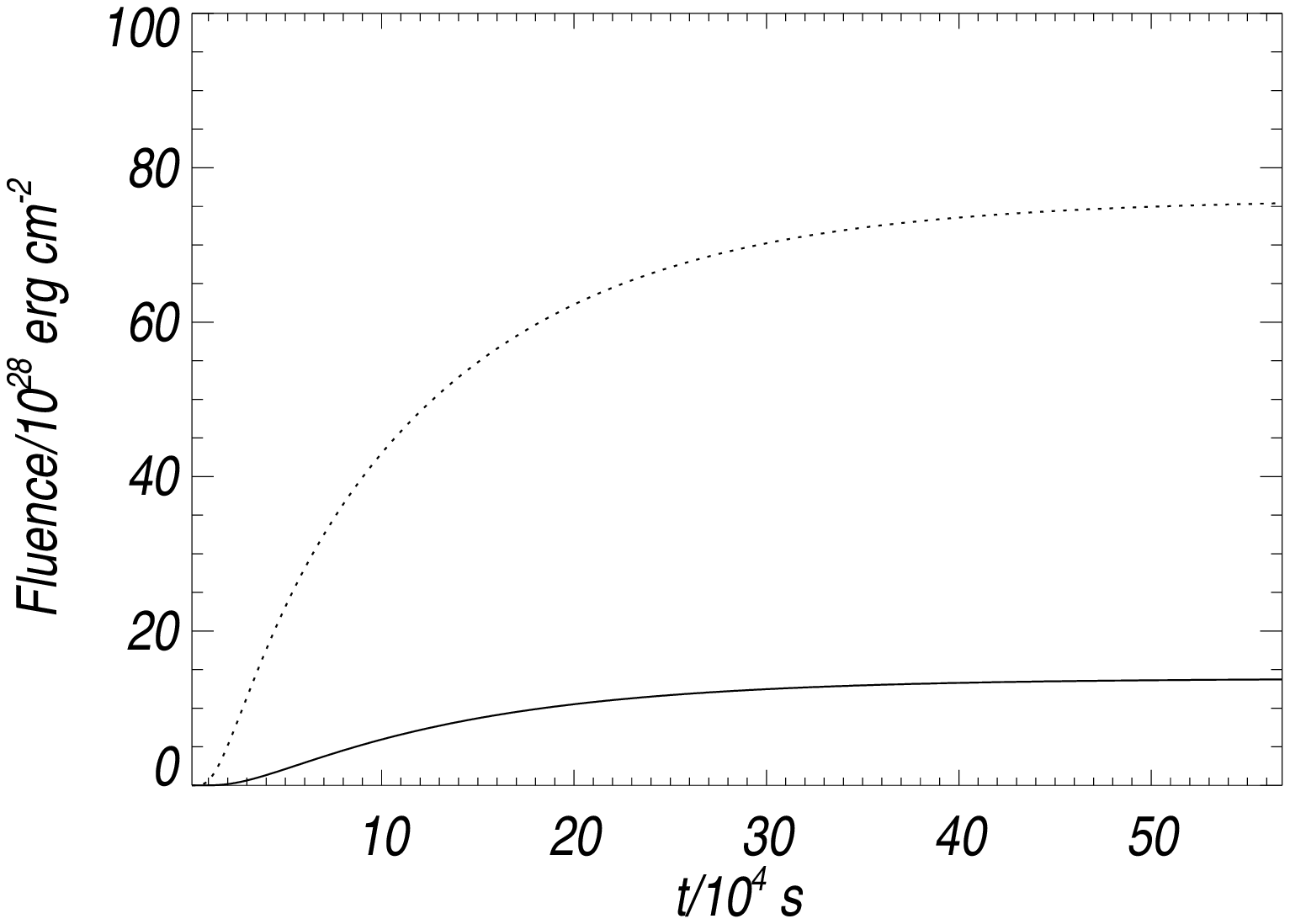]{The conductive fluence, in units of 
$10^{28}\usp\erg\usp\cm^{-2}$, for the model calculation described in 
eq.~[\protect\ref{eq:diffusion}], as a function of time in units of 
$10^{4}\usp\second$.  We show both the outward-directed (\emph{solid 
line}) and inward-directed (\emph{dotted line}) fluence.  As in 
Fig.~\protect\ref{fig:compare-fluence}, the evolution of the fluence is 
on the thermal decay timescale at the base of the burning layer.
\label{fig:fluence}}

\clearpage

\begin{figure}
\begin{center}
 \includegraphics[width=6in, height=4in]{f1.ps}
\end{center}

Figure 1: RXTE/ASM lightcurve of 4U 1820-30 prior to and around the epoch of 
the super burst. A flux of 1 Crab is approximately 75 ASM units.

\end{figure}
\clearpage

\begin{figure}
\begin{center}
 \includegraphics[width=6.5in, height=5in]{f2.ps}
\end{center}

Figure 2: Time history of the X-ray flux from 4U 1820-30 during the super 
burst. The PCA lightcurve (2 - 60 keV) at 1/8 s resolution is the higher time 
resolution trace (left axis). The lower time resolution curve is the 
(8 - 30)/(2 - 8) keV hardness ratio from Standard2 data with 16 s resolution 
(right axis).

\end{figure}
\clearpage

\begin{figure}
\begin{center}
 \includegraphics[width=6in, height=7in]{f3.ps}
\end{center}
\vskip 5pt

Figure 3: Comparison of the helium flash which
preceded the super burst (top panel) and a helium flash which was observed on 
May 2, 1997 (bottom panel). Shown for both bursts are the 2 - 60 keV 
lightcurves from Standard1 data with 1/8 s time resolution. The rates are
scaled to 3 PCU detectors.

\end{figure}
\clearpage

\begin{figure}
\begin{center}
 \includegraphics[width=4in, height=3in]{f4a.ps}
 \hspace{2in}%
\includegraphics[width=4in, height=3in]{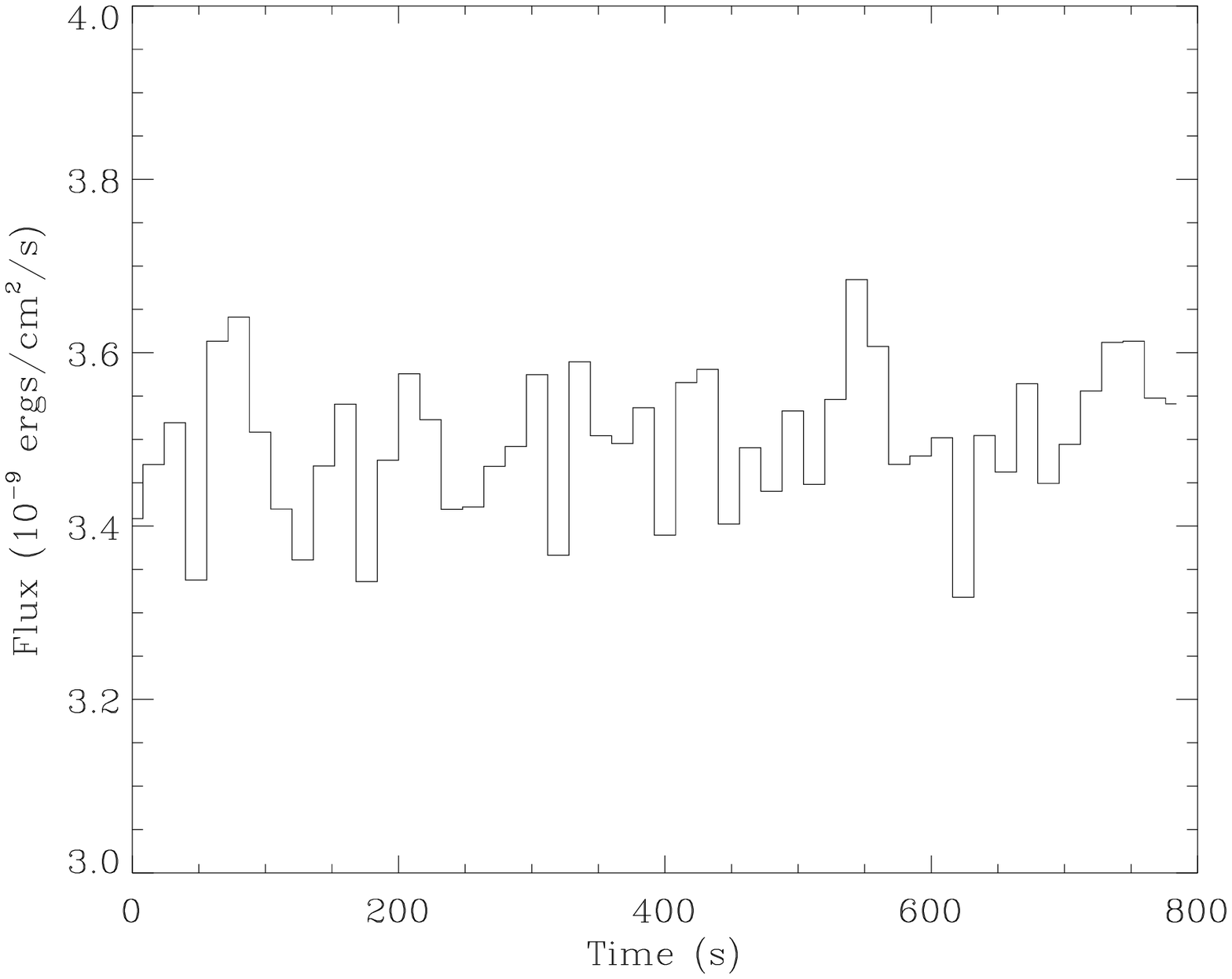}
\end{center}

Figure 4: Derived bremsstrahlung temperatures
(top panel) and 2 - 20 keV fluxes (bottom panel) for the persistent, accretion
driven flux prior to the super burst.

\end{figure}
\clearpage

\begin{figure}
\begin{center}
 \includegraphics[width=5in, height=4in]{f5.ps}
\end{center}

Figure 5: Residuals to a spectral fit during the 
peak of the burst using a black body function with photoelectric absorption.
Shown are the residuals, data - model, in units of standard deviations. The
emission line feature just longward of 6 keV as well as the edge between 
8 - 10 keV are cleary evident.

\end{figure}
\clearpage

\begin{figure}
\begin{center}
 \includegraphics[width=5in, height=6in]{f6.ps}
\end{center}

Figure 6: Count rate spectrum and model (top 
panel) and residuals, data - model in units of standard deviations 
(bottom panel) for a spectral fit to an interval during the peak of the burst. 
The model includes a black body function, a gaussian emission line near 6 kev
and an absorption edge near 9 keV. Note that the error bars near the peak of
the spectrum are almost smaller than the width of the lines.

\end{figure}
\clearpage

\begin{figure}
\begin{center}
 \includegraphics[width=6in, height=5in]{f7.ps}
\end{center}

Figure 7: Lightcurve of the burst from Standard1
data on a linear time axis. The dashed vertical lines denote the region in 
which we investigated the spectral evolution during the burst.

\end{figure}
\clearpage

\begin{figure}
\begin{center}
 \includegraphics[width=5in, height=7in]{f8.ps}
\end{center}
\vskip 10pt

Figure 8: Time evolution of the bolometric flux
(top panel), black body temperature (middle) and the inferred black body
radius (bottom) assuming a distance of 6.6 kpc. The time interval shown 
corresponds to the interval between the vertical dashed lines in figure 7.
The bolometric flux was derived by integrating the best-fit black body 
spectrum.

\end{figure}
\clearpage

\begin{figure}
\begin{center}
 \includegraphics[width=5in, height=7in]{f9.ps}
\end{center}
\vskip 20pt

Figure 9: Time evolution of the emission line and absorption edge
components throughout the burst.  Shown are the line centroid energy,
and line flux from the diskline model (top and 2nd from top) along
with the edge energy and depth (bottom and next to bottom) from the
smedge model.

\end{figure}
\clearpage

\begin{figure}
\begin{center}
 \includegraphics[width=5in]{f10.ps}
\end{center}
\vskip 20pt
    
Figure 10: The thermal structure of the other layers of a neutron star
accreting pure He at a local accretion rate $\dot{m}=2.25\times
10^4\usp\GramPerSc\usp\second^{-1}$.  The top axis indicates the time
required for a column $y$ to be accreted at this rate.  Plotted are the
temperature (\emph{solid line}) and flux (\emph{dotted line}) in units
of MeV per accreted nucleon.  The arrow is to guide the eye to the value
of $T$ at $y=10^{14}\usp\GramPerSc)$.  The dots indicates the base of
the C/Fe layer.  The condition for instability of the reaction
$\mathrm{^{12}C}+\mathrm{^{12}C}$ is indicated by the heavy dashed
curve.

\end{figure}
\clearpage

\begin{figure}
\begin{center}
 \includegraphics[width=5in]{f11.ps}
\end{center}
\vskip 20pt

Figure 11: The neutrino emissivity at a temperature of
$10^{10}\usp\K$, as function of column for a surface gravity
$2.34\ee{14}\usp\cm\usp\second^{-1}$.

\end{figure}
\clearpage

\begin{figure}
\begin{center}
 \includegraphics[width=5in]{f12.ps}
\end{center}
\vskip 20pt

Figure 12: The fluence, in units of $10^{28}\usp\erg\usp\cm^{-2}$, as
a function of time, in units of $10^{4}\usp\second$, for the one-zone
calculation (eq.~[\protect\ref{eq:one-zone-neutrino}]).  Shown are
both the neutrino (\emph{solid} line) and conductive (\emph{dotted}
line) fluences.  The conductive flux evolves on the thermal diffusion
timescale appropriate for a depth of $y=10^{13}\usp\GramPerSc$.

\end{figure}
\clearpage

\begin{figure}
\begin{center}
 \includegraphics[width=5in]{f13.ps}
\end{center}
\vskip 20pt

Figure 13: The luminosity, in units of
$10^{38}\usp\erg\usp\second^{-1}$, as a function of time, in units of
$10^{4}\usp\second$, for the one-zone calculation
(eq.~[\protect\ref{eq:one-zone-neutrino}]).

\end{figure}
\clearpage

\begin{figure}
\begin{center}
 \includegraphics[width=5in]{f14.ps}
\end{center}
\vskip 20pt

Figure 14: The conductive fluence, in units of
$10^{28}\usp\erg\usp\cm^{-2}$, for the model calculation described in
eq.~[\protect\ref{eq:diffusion}], as a function of time in units of
$10^{4}\usp\second$.  We show both the outward-directed (\emph{solid}
line) and inward-directed (\emph{dotted} line) fluence.  As in
Fig.~\protect\ref{fig:compare-fluence}, the evolution of the fluence
is on the thermal decay timescale at the base of the burning layer.

\end{figure}
\clearpage


\begin{thebibliography}{80}
\expandafter\ifx\csname natexlab\endcsname\relax\def\natexlab#1{#1}\fi

\bibitem[{{Anderson} {et~al.}(1997){Anderson}, {Margon}, {Deutsch}, {Downes},
  \& {Allen}}]{anderson.ea:timeresolved}
{Anderson}, S.~F., {Margon}, B., {Deutsch}, E.~W., {Downes}, R.~A., \& {Allen},
  R.~G. 1997, \apjl, 482, L69

\bibitem[{Arons \& King(1993)}]{arons.king:reprocessing}
Arons, J. \& King, I.~R. 1993, \apjl, 413, L121

\bibitem[{{Bildsten}(1995)}]{bildsten95:_propag}
{Bildsten}, L. 1995, \apj, 438, 852

\bibitem[{{Bildsten}(1998)}]{bildsten:thermonuclear}
{Bildsten}, L. 1998, in The Many Faces of Neutron Stars, ed. A.~Alpar,
  R.~Buccheri, \& J.~{van Paradijs}, Vol. 515, NATO ASI ser.~C (Dordrecht:
  Kluwer), 419

\bibitem[{Bildsten(2000)}]{bildsten.theory}
Bildsten, L. 2000, in Cosmic Explosions, ed. S.~S. Holt \& W.~W. Zhang

\bibitem[{{Bisnovatyi-Kogan} \&
  {Chechetkin}(1979)}]{bisnovatyi-kogan79:_noneq_x}
{Bisnovatyi-Kogan}, G.~S. \& {Chechetkin}, V.~M. 1979, Soviet Phys.-Uspekhi,
  127, 263

\bibitem[{{Blaes} {et~al.}(1990){Blaes}, {Blandford}, {Madau}, \&
  {Koonin}}]{blaes90:_slowl}
{Blaes}, O., {Blandford}, R., {Madau}, P., \& {Koonin}, S. 1990, \apj, 363, 612

\bibitem[{{Brown}(2000)}]{brown:nuclear}
{Brown}, E.~F. 2000, \apj, 531, 988

\bibitem[{{Brown} \& {Bildsten}(1998)}]{brown98}
{Brown}, E.~F. \& {Bildsten}, L. 1998, \apj, 496, 915

\bibitem[{{Brown} {et~al.}(1998){Brown}, {Bildsten}, \&
  {Rutledge}}]{brown98:transients}
{Brown}, E.~F., {Bildsten}, L., \& {Rutledge}, R.~E. 1998, \apjl, 504, L95

\bibitem[{{Buchler} \& {Yueh}(1976)}]{buchler.yueh:compton}
{Buchler}, J.~R. \& {Yueh}, W.~R. 1976, \apj, 210, 440

\bibitem[{Caughlan \& Fowler(1988)}]{caughlan88:_therm}
Caughlan, G.~R. \& Fowler, W.~A. 1988, At.\ Data Nucl.\ Data Tables, 40, 283

\bibitem[{{Clark} {et~al.}(1977){Clark}, {Li}, {Canizares}, {Hayakawa},
  {Jernigan}, \& {Lewin}}]{clark.ea:further}
{Clark}, G.~W., {Li}, F.~K., {Canizares}, C., {Hayakawa}, S., {Jernigan}, G.,
  \& {Lewin}, W. H.~G. 1977, \mnras, 179, 651

\bibitem[{Cornelisse {et~al.}(2000)Cornelisse, Heise, Kuulkers, Verbunt, \&
  in't Zand}]{cornelisse.ea:longest}
Cornelisse, R., Heise, J., Kuulkers, E., Verbunt, F., \& in't Zand, J. J.~M.
  2000, \apj, 357, L21

\bibitem[{Cumming \& Bildsten(2001)}]{cumming.bildsten:carbon}
Cumming, A. \& Bildsten, L. 2001, \apj, submitted, (astro-ph/0107213)

\bibitem[{{Day} \& {Done}(1991)}]{day.done:disc_reflected}
{Day}, C. S.~R. \& {Done}, C. 1991, \mnras, 253, 35P

\bibitem[{Ebisuzaki(1987)}]{ebisuzaki87}
Ebisuzaki, T. 1987, PASJ, 39, 287

\bibitem[{Ebisuzaki \& Nakamura(1988)}]{ebisuzaki.nakamura88}
Ebisuzaki, T. \& Nakamura, N. 1988, ApJ, 328, 251

\bibitem[{Farouki \& Hamaguchi(1993)}]{farouki93}
Farouki, R. \& Hamaguchi, S. 1993, \pre, 47, 4330

\bibitem[{{Foster} {et~al.}(1987){Foster}, {Fabian}, \&
  {Ross}}]{foster.ea:formation}
{Foster}, A.~J., {Fabian}, A.~C., \& {Ross}, R.~R. 1987, \mnras, 228, 259

\bibitem[{{Franco} \& {Strohmayer}(1999)}]{franco.strohmayer:detection}
{Franco}, L.~M. \& {Strohmayer}, T.~E. 1999, in American Astronomical Society
  Meeting, Vol. 195, 12609

\bibitem[{{Fryxell} \& {Woosley}(1982)}]{fryxell.woosley:two-dimensional}
{Fryxell}, B.~A. \& {Woosley}, S.~E. 1982, \apj, 258, 733

\bibitem[{Fushiki \& Lamb(1987)}]{fushiki87:_s}
Fushiki, I. \& Lamb, D.~Q. 1987, \apj, 317, 368

\bibitem[{{Grindlay} \& {Gursky}(1976)}]{grindlay.gursky:scattering}
{Grindlay}, J. \& {Gursky}, H. 1976, \apjl, 205, L131

\bibitem[{{Haberl} {et~al.}(1987){Haberl}, {Stella}, {White}, {Gottwald}, \&
  {Priedhorsky}}]{haberl.ea:exosat}
{Haberl}, F., {Stella}, L., {White}, N.~E., {Gottwald}, M., \& {Priedhorsky},
  W.~C. 1987, \apj, 314, 266

\bibitem[{{Haberl} \& {Titarchuk}(1995)}]{haberl.titarchuk}
Haberl, F. \& Titarchuk, L. 1995, A\&A, 299, 414

\bibitem[{{Haensel} \& {Zdunik}(1990)}]{haensel90a}
{Haensel}, P. \& {Zdunik}, J.~L. 1990, \aap, 227, 431

\bibitem[{Hansen \& Van~Horn(1975)}]{hansen75:_thin}
Hansen, C.~J. \& Van~Horn, H.~M. 1975, \apj, 195, 735

\bibitem[{{Heise} {et~al.}(2000){Heise}, {in't Zand}, \&
  {Kuulkers}}]{heise.ea:transient}
{Heise}, J., {in't Zand}, J. J.~M., \& {Kuulkers}, E. 2000, AAS/High Energy
  Astrophysics Division, 32, 2803+

\bibitem[{{Hesser} \& {Shawl}(1985)}]{hesser.shawl:integrated}
{Hesser}, J.~E. \& {Shawl}, S.~J. 1985, \pasp, 97, 465

\bibitem[{{Hoffman} {et~al.}(1978){Hoffman}, {Lewin}, {Doty}, {Jernigan},
  {Haney}, \& {Richardson}}]{hoffman.ea:sas_observation}
{Hoffman}, J.~A., {Lewin}, W. H.~G., {Doty}, J., {Jernigan}, J.~G., {Haney},
  M., \& {Richardson}, J.~A. 1978, \apjl, 221, L57

\bibitem[{{Inogamov} \& {Sunyaev}(1999)}]{inogamov.sunyaev:spread}
{Inogamov}, N.~A. \& {Sunyaev}, R.~A. 1999, Astronomy Letters, 25, 269

\bibitem[{Itoh {et~al.}(1996)Itoh, Hayashi, Nishikawa, \&
  Kohyama}]{itoh96:_neutr}
Itoh, N., Hayashi, H., Nishikawa, A., \& Kohyama, Y. 1996, \apjs, 102, 411

\bibitem[{{Itoh} {et~al.}(1991){Itoh}, {Kuwashima}, {Ichihashi}, \&
  {Mutoh}}]{itoh91:_rossel_gaunt}
{Itoh}, N., {Kuwashima}, F., {Ichihashi}, K., \& {Mutoh}, H. 1991, \apj, 382,
  636

\bibitem[{{Joss} \& {Li}(1980)}]{joss80:_helium}
{Joss}, P.~C. \& {Li}, F.~K. 1980, \apj, 238, 287

\bibitem[{{Kaaret} {et~al.}(1997){Kaaret}, {Ford}, \&
  {Chen}}]{kaaret97:_stron_field_gener_relat_quasi}
{Kaaret}, P., {Ford}, E.~C., \& {Chen}, K. 1997, \apjl, 480, L27

\bibitem[{{Kuulkers}(2001)}]{kuulkers:superoutburst}
{Kuulkers}, E. 2001, Ast.\ Tel., 68, 1

\bibitem[{{Lewin} {et~al.}(1984){Lewin}, {Vacca}, \&
  {Basinska}}]{lewin.ea:precursors}
{Lewin}, W. H.~G., {Vacca}, W.~D., \& {Basinska}, E.~M. 1984, \apjl, 277, L57

\bibitem[{Li {et~al.}(2001){Li}, {Gu}, \& {Kahn}}]{Li.ea01}
Li, Y., Gu, M. F. \& Kahn, S. M. 2001, ApJ, submitted, 
(astro-ph/0106163)

\bibitem[{Litwin {et~al.}(2001)Litwin, Brown, \&
  Rosner}]{litwin.brown.ea:ballooning}
Litwin, C., Brown, E.~F., \& Rosner, R. 2001, \apj, 533, 788

\bibitem[{London {et~al.}(1986)London, Taam, \& Howard}]{london.ea.86}
London, R. A., Taam, R. E., \& Howard, W. M. 1986, ApJ, 306, 170

\bibitem[{Madej(1997)}]{madej97}
Madej, J. 1997, A\&A, 320, 177

\bibitem[{Madej(1991)}]{madej91}
Madej, J. 1991, ApJ, 376, 161

\bibitem[{{Madej}(1989)}]{madej:Compton}
{Madej}, J. 1989, \apj, 339, 386

\bibitem[{Madsen {et~al.}(1979)Madsen, , \& Sincovec}]{madsen.sincovec:pdecol}
Madsen, N.~K., , \& Sincovec, R.~F. 1979, ACM Trans.~Math.~Soft., 5, 326

\bibitem[{{Magnier} {et~al.}(1989){Magnier}, {Lewin}, {van Paradijs}, {Tan},
  {Penninx}, \& {Damen}}]{magnier.ea:spectral}
{Magnier}, E., {Lewin}, W. H.~G., {van Paradijs}, J., {Tan}, J., {Penninx}, W.,
  \& {Damen}, E. 1989, \mnras, 237, 729

\bibitem[{{Nakamura} {et~al.}(1988){Nakamura}, {Inoue}, \&
  {Tanaka}}]{nakamura.ea:detection}
{Nakamura}, N., {Inoue}, H., \& {Tanaka}, Y. 1988, \pasj, 40, 209

\bibitem[{Nayakshin \& Kallman(2001)}]{nayakshin.kallman.01}
Nayakshin, S., Kallman, T. R. 2001, ApJ, 546, 406

\bibitem[{Ogata {et~al.}(1993)Ogata, Ichimaru, \& {van Horn}}]{ogata93:_therm}
Ogata, S., Ichimaru, S., \& {van Horn}, H.~M. 1993, \apj, 417, 265

\bibitem[{Popham \& Sunyaev(2001)}]{popham.sunyaev:accretion}
Popham, R. \& Sunyaev, R. 2001, \apj, 547, 355

\bibitem[{{Potekhin} {et~al.}(1999){Potekhin}, {Baiko}, {Haensel}, \&
  {Yakovlev}}]{potekhin99:_trans}
{Potekhin}, A.~Y., {Baiko}, D.~A., {Haensel}, P., \& {Yakovlev}, D.~G. 1999,
  \aap, 346, 345

\bibitem[{{Potekhin} {et~al.}(1997){Potekhin}, {Chabrier}, \&
  {Yakovlev}}]{potekhin97}
{Potekhin}, A.~Y., {Chabrier}, G., \& {Yakovlev}, D.~G. 1997, \aap, 323, 415

\bibitem[{{Priedhorsky} \& {Terrell}(1984)}]{priedhorsky.terrell:discovery}
{Priedhorsky}, W. \& {Terrell}, J. 1984, \apjl, 284, L17

\bibitem[{{Rappaport} {et~al.}(1987){Rappaport}, {Ma}, {Joss}, \&
  {Nelson}}]{rappaport.ea:evolutionary}
{Rappaport}, S., {Ma}, C.~P., {Joss}, P.~C., \& {Nelson}, L.~A. 1987, \apj,
  322, 842

\bibitem[{{Rich} {et~al.}(1993){Rich}, {Minniti}, \&
  {Liebert}}]{rich.ea:far_uv}
{Rich}, R.~M., {Minniti}, D., \& {Liebert}, J. 1993, \apj, 406, 489

\bibitem[{Ross {et~al.}(1999)Ross, Fabian, \& Young}]{ross.ea.99}
Ross, R. R., Fabian, A. C. \& Young, A. J. 1999, MNRAS, 306, 461

\bibitem[{{Sampson}(1959)}]{sampson:opacity}
{Sampson}, D.~H. 1959, \apj, 129, 734

\bibitem[{{Schatz} {et~al.}(2001){Schatz}, {Aprahamian}, {Barnard}, {Bildsten},
  {Cumming}, {Ouellette}, {Rauscher}, {Thielemann}, \&
  {Wiescher}}]{schatz.aprahamian.ea:endpoint}
{Schatz}, H., {Aprahamian}, A., {Barnard}, V., {Bildsten}, L., {Cumming}, A.,
  {Ouellette}, M., {Rauscher}, T., {Thielemann}, F.-K., \& {Wiescher}, M. 2001,
  \prl, in press

\bibitem[{{Schatz} {et~al.}(1999){Schatz}, {Bildsten}, {Cumming}, \&
  {Wiescher}}]{schatz99}
{Schatz}, H., {Bildsten}, L., {Cumming}, A., \& {Wiescher}, M. 1999, \apj, 524,
  1014
  
\bibitem[{{Smale} {et~al.}(1997){Smale}, {Zhang}, \&
  {White}}]{smale.ea:discovery}
{Smale}, A.~P., {Zhang}, W., \& {White}, N.~E. 1997, \apjl, 483, L119

\bibitem[{{Spitkovsky} {et~al.}(2001){Spitkovsky}, {Levin}, and
{Ushomirsky}}]{spitkovsky.ea.propagation}
{Spitkovsky}, A., {Levin}, Y., \& {Ushomirsky}, G. 2001, \apj,
submitted.  preprint available: astro-ph/0108074

\bibitem[{{Stella} {et~al.}(1987){Stella}, {White}, \&
  {Priedhorsky}}]{stella.ea:discovery}
{Stella}, L., {White}, N.~E., \& {Priedhorsky}, W. 1987, \apjl, 312, L17

\bibitem[{{Strohmayer}(2000)}]{strohmayer:remarkable}
{Strohmayer}, T.~E. 2000, AAS/High Energy Astrophysics Division, 32, 2410

\bibitem[{{Taam} \& {Picklum}(1978)}]{taam78:_nuclear}
{Taam}, R.~E. \& {Picklum}, R.~E. 1978, \apj, 224, 210

\bibitem[{{Tawara} {et~al.}(1984){Tawara}, {Kii}, {Hayakawa}, {Kunieda},
  {Masai}, {Nagase}, {Inoue}, {Koyama}, {Makino}, {Makishima}, {Matsuoka},
  {Murakami}, {Oda}, {Ogawara}, {Ohashi}, {Shibazaki}, {Tanaka}, {Miyamoto},
  {Tsunemi}, {Yamashita}, \& {Kondo}}]{tawara.ea:very_long}
{Tawara}, Y., {Kii}, T., {Hayakawa}, S., {Kunieda}, H., {Masai}, K., {Nagase},
  F., {Inoue}, H., {Koyama}, K., {Makino}, F., {Makishima}, K., {Matsuoka}, M.,
  {Murakami}, T., {Oda}, M., {Ogawara}, Y., {Ohashi}, T., {Shibazaki}, N.,
  {Tanaka}, Y., {Miyamoto}, S., {Tsunemi}, H., {Yamashita}, K., \& {Kondo}, I.
  1984, \apjl, 276, L41

\bibitem[{{Timmes} \& {Swesty}(2000)}]{timmes.swesty:accuracy}
{Timmes}, F.~X. \& {Swesty}, F.~D. 2000, \apjs, 126, 501

\bibitem[{{Timmes} \& {Woosley}(1992)}]{timmes92}
{Timmes}, F.~X. \& {Woosley}, S.~E. 1992, \apj, 396, 649

\bibitem[{Titarchuk(1994)}]{titarchuk.94}
Titarchuk, L. 1994, ApJ, 429, 340

\bibitem[{{Townsend} (1966)}]{townsend:entrainment}
{Townsend}, A.~A. 1966, J. Fluid Mech., 1, 29

\bibitem[{Turner {et~al.}(1992)}]{turner.ea.92}
Turner, T. J., Done, C., Mushotzky, R., Madejski, G. \& Kunieda, H.
1992, ApJ, 391, 102

\bibitem[{Urpin \& Yakovlev(1980)}]{urpin80:_therm}
Urpin, V.~A. \& Yakovlev, D.~G. 1980, \sovast, 24, 126

\bibitem[{{Vacca} {et~al.}(1986){Vacca}, {Lewin}, \& {van
  Paradijs}}]{vacca.ea:eddington}
{Vacca}, W.~D., {Lewin}, W. H.~G., \& {van Paradijs}, J. 1986, \mnras, 220, 339

\bibitem[{{van Paradijs} {et~al.}(1990){van Paradijs}, {Dotani}, {Tanaka}, \&
  {Tsuru}}]{vanparadijs.ea:very_energetic}
{van Paradijs}, J., {Dotani}, T., {Tanaka}, Y., \& {Tsuru}, T. 1990, \pasj, 42,
  633

\bibitem[{{Waki} {et~al.}(1984){Waki}, {Inoue}, {Koyama}, {Matsuoka},
  {Murakami}, {Ogawara}, {Ohashi}, {Tanaka}, {Hayakawa}, {Tawara}, {Miyamoto},
  {Tsunemi}, \& {Kondo}}]{waki.ea:discovery}
{Waki}, I., {Inoue}, H., {Koyama}, K., {Matsuoka}, M., {Murakami}, T.,
  {Ogawara}, Y., {Ohashi}, T., {Tanaka}, Y., {Hayakawa}, S., {Tawara}, Y.,
  {Miyamoto}, S., {Tsunemi}, H., \& {Kondo}, I. 1984, \pasj, 36, 819

\bibitem[{{Walker} \& {Meszaros}(1989)}]{walker.meszaros:dynamical}
{Walker}, M.~A. \& {Meszaros}, P. 1989, \apj, 346, 844

\bibitem[{{Wijnands}(2001)}]{wijnands:recurrent}
{Wijnands}, R. 2001, \apjl, 554, L59

\bibitem[{{Woosley} \& {Taam}(1976)}]{woosley.taam:carbon}
{Woosley}, S.~E. \& {Taam}, R.~E. 1976, \nat, 263, 101

\bibitem[{Youngs(1994)}]{youngs:RT}
Youngs, D.~L. 1994, Lasers and Particle Beams, 12, 725

\bibitem[{{Zhang} {et~al.}(1997){Zhang}, {Strohmayer}, \&
  {Swank}}]{zhang.ea:neutron_star_masses}
{Zhang}, W., {Strohmayer}, T.~E., \& {Swank}, J.~H. 1997, \apjl, 482, L167

\bibitem[{{Zingale} {et~al.}(2001){Zingale}, {Timmes}, {Fryxell}, {Lamb},
  {Olson}, {Calder}, {Dursi}, {Ricker}, {Rosner}, {MacNeice}, \&
  {Tufo}}]{zingale.timmes.ea:helium}
{Zingale}, M., {Timmes}, F.~X., {Fryxell}, B., {Lamb}, D.~Q., {Olson}, K.,
  {Calder}, A.~C., {Dursi}, L.~J., {Ricker}, P., {Rosner}, R., {MacNeice}, P.,
  \& {Tufo}, H.~M. 2001, \apjs, in press

\end{thebibliography}
\end{document}